\documentclass[sigconf]{acmart}

\AtBeginDocument{%
  }

\copyrightyear{2026}
\acmYear{2026}
\setcopyright{cc}
\setcctype{by}
\acmConference[CHI '26]{Proceedings of the 2026 CHI Conference on Human Factors in Computing Systems}{April 13--17, 2026}{Barcelona, Spain}
\acmBooktitle{Proceedings of the 2026 CHI Conference on Human Factors in Computing Systems (CHI '26), April 13--17, 2026, Barcelona, Spain}
\acmPrice{}
\acmDOI{10.1145/3772318.3790475}
\acmISBN{979-8-4007-2278-3/2026/04}

\usepackage{graphicx}
\usepackage{multirow}
\usepackage{subcaption}
\usepackage{cleveref}
\usepackage{array}
\usepackage{enumitem}
\usepackage{soul}

\begin{document}

\title[Counting the Wait]{Counting the Wait: Effects of Temporal Feedback on Downstream Task Performance and Perceived Wait-Time Experience during System-Imposed Delays}


\author{Felicia Fang-Yi Tan}
\email{felicia.tan@nyu.edu}
\orcid{0000-0001-7837-174X}
\affiliation{%
  \institution{Tandon School of Engineering}
  \institution{New York University}
  \city{New York}
  \country{USA}
}

\author{Oded Nov}
\email{onov@nyu.edu}
\orcid{0000-0001-6410-2995}
\affiliation{%
  \institution{Tandon School of Engineering}
  \institution{New York University}
  \city{New York}
  \country{USA}
}


\renewcommand{\shortauthors}{Tan and Nov}

\begin{abstract}

System-imposed wait times can significantly disrupt digital workflows, affecting user experience and task performance. Prior HCI research has examined how temporal feedback, such as feedback mode (Elapsed-Time vs. Remaining-Time) shapes wait-time perception. However, few studies have investigated how such feedback influences users’ downstream task performance, as well as overall affective and cognitive experience. To study these effects, we conducted an online experiment where 425 participants performing a visual reasoning task experienced a 10-, 30-, or 60-second wait with a Remaining-Time, Elapsed-Time, or No Time Display. Findings show that temporal feedback mode shapes how waiting is perceived: Remaining-Time feedback increased frustration relative to Elapsed-Time feedback, while No Time Display made waits feel longer and heightened ambiguity. Notably, these experiential differences did not translate into differences in post-wait task performance. Integrating psychophysical and cognitive science perspectives, we discuss implications for implementing temporal feedback in latency-prone digital systems.

\end{abstract}





% \vspace{-0.6cm}
%     \centering
%     \includegraphics[width=\textwidth]{Figures/teaser.png}
%     \vspace{-0.7cm}
%     \caption{XX}
%   \Description{XX}
%   \vspace{-0.2cm}

%%\received{20 February 2007}
%%\received[revised]{12 March 2009}
%%\received[accepted]{5 June 2009}

\maketitle

\section{Introduction}

Wait times caused by system delays permeate digital interactions, from loading web content and buffering media streams to generating outputs from computationally intensive AI models \cite{ghafurian_countdown_2020, arapakis_impact_2014}.  Prior research has established that even modest increases in system delay lead to higher abandonment rates, reduced task efficiency, and poorer perceptions of system quality \cite{boucsein_forty_2009, galletta_when_2006, hoxmeier_system_2000}. For example, users typically tolerate only about two seconds of delay without feedback in information retrieval tasks, with abandonment rates sharply rising after approximately 10-15 seconds \cite{nah_study_2004, selvidge_how_1999}. While engineering-level optimizations continue striving to minimize latency (i.e., the delay between user input and system response \cite{attig_system_2017}), residual wait times remain inevitable. Consequently, the burden of managing delays has increasingly shifted towards effective interface design, making the design of temporal feedback a central concern in HCI.

Prior research on temporal feedback has produced valuable guidelines around latency thresholds and interface design strategies \cite{dabrowski_40years_2011, attig_system_2017}. Studies comparing count-down versus count-up indicators have shown that displays emphasizing remaining time often feel shorter and more reassuring than displays showing time elapsed \cite{shalev_does_2013, komatsu_waiting_2024}. These effects can be understood through psychophysical principles. The filled-duration illusion describes how visual dynamics influence perceived duration \cite{gomez_filled-duration_1979}, while Weber’s Law clarifies perceptual thresholds for noticing changes in duration \cite{haigh_role_2021}. However, prior experiments (see Section~\ref{RelatedWork - HCI and Human‐Factors}) predominantly fall into two categories: (i) time-perception studies that ask participants to watch a timer or neutral content without a concurrent primary task (e.g., \cite{hong_when_2013, shalev_does_2013, komatsu_waiting_2024, ghafurian_countdown_2020}), and (ii) studies that embed temporal feedback in an interactive task but collect immediate post-delay judgments (e.g., \cite{hurter_active_2012, galletta_when_2006}), before any task resumption. In both cases, what happens next is overlooked: how temporal feedback during the wait shapes users’ ability to re-engage with an interrupted task. A distinct stream of interruptions literature does measure post-interruption resumption and performance, but it typically manipulates when and how interruptions occur (e.g., notifications, breakpoints) rather than exploring the effects of temporal feedback during a system-imposed wait. Consequently, we know little about the downstream impact of temporal feedback on post-wait performance.

Our study (n=425) addresses this gap with an online experiment, where participants resume an ongoing visual reasoning task after experiencing a system-imposed wait. Instead of comparing count-up and count-down timers as two variants of directional display shown during the wait time, we adopt a framing that distinguishes feedback by the type of temporal information it provides. Specifically, our framing organizes temporal feedback into three informational modes: (i) withholding temporal information during the wait (\emph{No Time Display}); (ii) conveying time already passed (\emph{Elapsed-Time}); and (iii) communicating time remaining until completion (\emph{Remaining-Time}).

We address the following research questions:

\begin{enumerate}[label=(RQ\arabic*)]
    \item How do temporal feedback mode and wait duration interact to affect subsequent task performance in interrupted workflows?  
    \item How do temporal feedback mode and wait duration influence emotional and cognitive responses (i.e., affect, workload, time perception) throughout the overall experience?
    \item How do participants perceive the influence of the wait on their overall experience?  
\end{enumerate}

The findings show that temporal feedback mode and wait duration shaped users’ subjective experience. \emph{Remaining-Time} significantly increased frustration and reduced pleasantness ratings compared to \emph{Elapsed-Time}, and \emph{No Time Display} made waits feel subjectively longer than \emph{Remaining-Time}. Longer waits were associated with more opportunities for mental recovery. None of these experiential differences, however, translated into significant differences in post-wait task performance. These results extend existing work by showing that temporal feedback can meaningfully alter the affective and cognitive experience of waiting without producing corresponding changes in downstream task outcome. This experiential–performance dissociation adds a new perspective to research on waiting and interruptions, which often emphasizes the benefits of temporal cues for supporting resumption. In our context, temporal feedback shaped how the wait was experienced rather than how users performed after the wait. We interpret this dissociation as evidence that experiential responses and performance outcomes may be influenced by different underlying mechanisms. Notably, while our task affords controlled measurement of post-wait performance within an ongoing activity, its simplified structure limits ecological breadth. Future work should test whether these dissociations persist in more complex workflows such as writing, collaborative editing, or multitasking systems.

Accordingly, our work contributes:

\begin{enumerate}
    \item Empirical insights into how temporal feedback modes and wait durations jointly shape \emph{downstream} task performance, subjective time perception, cognitive workload, and emotional response.
    \item Theory-informed design guidelines combining psychophysical and cognitive workload principles with empirical findings, providing actionable recommendations for designing temporal feedback in latency-prone digital systems.
\end{enumerate}

\section{Related Work}

We organize the review of prior work into three strands that together frame our study: (i) \emph{Temporal Feedback in HCI} - research on temporal feedback (progress bars, timers) and latency thresholds; (ii) \emph{Interruptions and task resumption} - studies on when/how breaks occur and their resumption costs; and (iii) \emph{Psychophysics of time perception} - mechanisms explaining why interface cues reshape perceived duration and affect.

\subsection{Temporal Feedback in HCI}
\label{RelatedWork - HCI and Human‐Factors}

Decades of HCI work specify latency thresholds and recommend visual feedback during waits (progress bars, spinners, timers) to reduce abandonment and frustration by increasing predictability and perceived control \cite{boucsein_forty_2009, dabrowski_40years_2011, myers_importance_1985}. Building on this foundation, prior research can be organized around three design levers: whether to show time at all, what temporal information to present, and how to render it.

First, for studies on \emph{whether} to show time, explicit feedback generally improves perceived reliability and reduces uncertainty relative to showing nothing \cite{lallemand_enhancing_2012, dellaert_how_1999}. However, directing attention to time can also sustain vigilance and make waits \emph{feel} longer in some contexts, especially as durations increase \cite{lallemand_enhancing_2012}. Recent work comparing implicit versus explicit feedback similarly shows that highly informative cues boost perceived control and attention, but may not always shorten perceived duration; under certain settings, less explicit cues can \emph{feel} faster \cite{fang_effects_2022}. In short, showing \emph{No Time Display} can dampen time monitoring on brief waits but trades off predictability as delays lengthen \cite{hong_when_2013, fang_effects_2022}.

Second, for work on \emph{what} to show, prior research shows that directionality matters. \emph{Remaining-Time} indicators (count-downs) often compress subjective time more than \emph{Elapsed-Time} displays (count-ups) and can yield more positive downstream evaluations in lab tasks \cite{shalev_does_2013}. That advantage, however, depends on details such as count range (e.g., the effect is strongest when a count-down runs toward a salient endpoint such as 1 or 0) and context \cite{shalev_does_2013}. Beyond generic interfaces, applied domains such as traffic signals show that communicating remaining time reduces uncertainty and supports readiness for action \cite{cui_feedback_2022}.

Third, for research on \emph{how} to render feedback, salience, pacing, and update dynamics were found to shape perceived duration and affect. Classic and contemporary studies show that dynamic, attention-capturing cues alter time judgments (e.g., non-linear progress, visual “filling”), consistent with filled-duration and Weber-scaling accounts \cite{gomez_filled-duration_1979, haigh_role_2021}. Manipulating progress-bar kinematics (e.g., ribbing, non-linear speed, or end-weighting) can make identical waits \emph{appear} faster \cite{harrison_rethinking_2007, harrison_faster_2010}. Other work cautions that high salience and rapid updates can elevate arousal during long waits, while subtler pacing can reduce time monitoring \cite{myers_importance_1985, fang_effects_2022}. Recent findings show that tempo engineering (varying the timer’s tick rate while keeping the nominal duration fixed) and chunking (showing fewer, coarser updates or discrete milestones) alter sensitivity to discrepancies: faster early ticks followed by slower ticks can compress perceived time and increase detection of irregularities \cite{komatsu_waiting_2024}, whereas coarser updates and moderate tick rates can mask small deviations and sometimes shorten perceived waits \cite{cui_feedback_2022, ghafurian_countdown_2020}. Designers have also explored augmenting waits with lightweight micro-activities (e.g., “active progress bars”) to repurpose idle time and improve satisfaction without harming primary-task return \cite{hurter_active_2012, harrison_rethinking_2007}.

Despite the extant research, a methodological gap persists which limits the applicability of prior research to the understanding of behavior and perception associated with wait times. Many studies measure perceptions or satisfaction immediately after a wait, with the wait as the terminal event \cite{lallemand_enhancing_2012, dellaert_how_1999, fang_effects_2022}, rather than examining how feedback choices affect \emph{post-wait task resumption} once users continue an interrupted task. Emerging application-specific work on AI content generation echoes these themes. Users infer quality and manage expectations from generation cues during waits, yet this work similarly centers on in-wait perceptions rather than downstream performance \cite{lee_while_2025}.

Overall, existing research shows that explicit temporal feedback, especially \emph{Remaining-Time}, often reassures users and can compress perceived waiting time, but leaves open how these design choices shape cognitive workload, affect, and task performance \emph{after} the wait. Our work addresses this gap by embedding either \emph{No Time Display}, \emph{Elapsed-Time}, or \emph{Remaining-Time} into an ongoing task and evaluating both in-wait experience and downstream outcomes across 10-, 30-, 60-second delays.

\subsection{Interruptions, Embedded Waits, and Task Resumption}
\label{Related Work-interruptions}

In our context, a system-imposed wait functions as a brief interruption of an ongoing task, and decades of work show that such suspensions impose resumption costs (such as slower performance, more errors, and higher workload), especially when they arrive at inopportune moments \cite{altmann_memory_2002, monk_effect_2008}. In such cases, timing policies matter. Aligning interruptions with natural task boundaries reliably reduces disruption relative to mid-subtask moments, while non-aligned or random timing elevates perceived workload \cite{adamczyk_if_2004, bailey_need_2006}. Beyond timing, even short machine-imposed delays can shift strategy from largely automatic execution to step-by-step monitoring and verification, indicating greater cognitive control demands under delay \cite{odonnell_how_1996}. These costs are observable in the wild - in web search, users notice added response time on the order of \(\sim\)1\,s and preferentially engage with lower-latency but content-identical result pages \cite{arapakis_impact_2014}. Likewise, small per-operation delays act as embedded micro-waits within an ongoing task. In exploratory visual analysis, adding \(\sim\)500\,ms to each interaction reduced interaction volume and dataset coverage and lowered the rate of observations, generalizations, and hypotheses, with dampened performance persisting even after latency was removed \cite{liu_effects_2014}. 

Goal-activation accounts explain why interruptions negatively affect the knowledge worker. Suspending a goal induces decay and interference, so resumption requires cue-based reinstatement; brief rehearsal during the interruption and salient resumption cues reduce lag \cite{altmann_memory_2002, trafton_task_2007}. Longer and more demanding interludes exacerbate costs via decay and resource competition, consistent with multiple-resource theory \cite{wickens_multiple_2008}. 

Critically for our study, this literature typically manipulates \emph{when} or \emph{how} interruptions occur (e.g., breakpoints, notifications, secondary tasks) rather than the temporal feedback provided during the forced wait. As a result, we know little about whether showing \emph{Remaining-Time}, \emph{Elapsed-Time}, or \emph{No Time Display} during the pause mitigates or amplifies downstream resumption costs on speed, accuracy, workload, and affect.

\subsection{Psychophysics of Waiting and Time Perception}
\label{sec:related-psychophysics}

Design choices for temporal feedback do more than inform users - they also redirect attention and reshape the perceived length of a wait. Classic psychophysics explains why small temporal cues in the interface can stretch or compress subjective time and color affect.

A foundational account is the \emph{pacemaker–accumulator} model: an internal pacemaker emits pulses that are integrated to estimate duration, and attention gates those pulses. Directing attention to time opens the gate so more pulses reach the accumulator (longer felt time); engaging attention elsewhere narrows the gate (shorter felt time) \cite{treisman_temporal_1963, gibbon_scalar_1984, zakay_role_1996}. External temporal cues-on-screen clocks and progress timers-hold the gate open and increase estimation precision by keeping time top-of-mind \cite{reed_use_2004}. Physiological arousal also matters: higher arousal (e.g., stress, exertion) speeds the pacemaker and inflates experienced duration \cite{droit-volet_emotional_2016, ozel_effect_2004, boltz_changes_1994}. In HCI terms, a salient countdown invites temporal attention (gate open), whereas rich placeholder content or micro-tasks draw attention off the clock (gate narrowed).

Two well-established phenomena further clarify design impacts. First, the \emph{filled-duration illusion} explains that intervals packed with changes (e.g., movement, flashing, incremental updates) are judged longer than “empty” intervals of equal length \cite{gomez_filled-duration_1979}. Thus, highly animated progress may paradoxically lengthen the felt wait even as it reassures users that work is ongoing. Second, Weber’s law implies that just-noticeable differences in duration scale with the base interval, i.e., small overruns are more salient (and frustrating) on short waits than on long ones \cite{haigh_role_2021}. Together, these principles explain why minimalist cues can make brief waits feel brisk and less intrusive, while the same cues can drag during longer operations.

Additionally, a related distinction is \emph{prospective} versus \emph{retrospective} timing. When users know in advance they will judge time (prospective), or are repeatedly shown time cues, they allocate more attention to time. When they do not, remembered duration (retrospective) leans more on contextual change. Meta-analytic evidence shows that prospective judgments are typically longer and less variable than retrospective ones, and that task difficulty (which uses attention) selectively shortens prospective estimates \cite{block_prospective_1997}. These findings align with attentional-gate predictions and help interpret why the presence of temporal feedback can elongate the felt wait, whereas drawing attention into content or micro-interactions (e.g., showing skeleton previews of the next items to read, or letting the user rename an uploaded file) can compress it.

\section{Study}

Building on past HCI and psychophysical research, we designed a controlled experiment manipulating two factors central to waiting experiences in digital workflows:  \textbf{Temporal Feedback Mode} (\emph{No Time Display}, \emph{Elapsed-Time}, \emph{Remaining-Time}) and \textbf{Wait Duration} (10, 30, 60 seconds). The next sections describe the motivations, design choices, and operational details of each factor. We evaluated post-wait task performance on the portion of the task completed after the system-imposed delay, and collected cognitive workload, affect, and time-perception ratings at the end of the full task session.

To examine these effects, participants completed a visual reasoning task divided into two segments, with a system-imposed wait inserted between them to simulate an interruption that occurs during screen-based tasks. This allowed us to address the three research questions: \textbf{RQ1}, how temporal feedback mode and wait duration interact to affect subsequent task performance in interrupted workflows; \textbf{RQ2}, how these factors influence emotional and cognitive responses (i.e., affect, workload, time perception) of the overall session experience; and \textbf{RQ3}, how participants interpret and describe the influence of the wait on their overall experience. We examined these outcomes using quantitative and qualitative measures detailed below.

\subsection{Participants}
    
    \begin{table}[H]
    \caption{Participant demographics. Group-wise balance checks were conducted (ANOVA for age; chi-square tests for sex and ethnicity), confirming no demographic differences across the 3$\times$3 experimental conditions.}
    \Description{Participant demographics (n=425). Sex: Female 209, Male 214, Prefer not to say 2. Age: Majority 25–54 (308 total); remaining - 18–24: 39; 55–64: 52; 65+: 26. Ethnicity: White 306; Black 48; Asian 28; Mixed 24; Other 17; Unknown 2. Group-wise balance checks were conducted (ANOVA for age; chi-square tests for sex and ethnicity), confirming no demographic differences across the 3×3 experimental conditions.}
    \label{tab:demographics}
    \centering
    %\small
    \setlength{\tabcolsep}{4pt}
    \renewcommand{\arraystretch}{0.9}
    \setlength{\tabcolsep}{3.5pt}
    \begin{tabular}{l r | l r | l r}
    \toprule
    \textbf{Sex} & \textbf{Count} &
    \textbf{Age} & \textbf{Count} &
    \textbf{Ethnicity} & \textbf{Count} \\
    \midrule
    Female & 209 & 18--24 & 39 & White & 306 \\
    Male & 214 & 25--34 & 110 & Black & 48 \\
    Prefer not to say & 2 & 35--44 & 101 & Asian & 28 \\
    & & 45--54 & 97 & Mixed & 24 \\
    & & 55--64 & 52 & Other & 17 \\
    & & 65+ & 26 & Unknown & 2 \\
    \bottomrule
    \end{tabular}
    \end{table}

We recruited participants through the Prolific crowdsourcing platform\footnote{\url{https://www.prolific.com}}. Eligibility was restricted to individuals residing in the United States who had completed at least 100 prior studies on the platform with a minimum approval rating of 95\%. Because the task required distinguishing object colors, participation was limited to individuals who reported no color-vision deficiencies, as specified in the Prolific screening and consent materials.

An a priori power analysis for a 3×3 between-subjects ANOVA (f = 0.25, $\alpha$ = 0.05, power = 0.80) suggested a minimum sample size of N = 249. To account for potential exclusions and ensure sufficient power for interaction effects, we recruited 600 participants initially. 19 participants were excluded for failing an attention check, and 165 participants were excluded for completing less than 80\% of the task (9 participants met both exclusion criteria). The 80\% completion-rate threshold was applied to ensure sufficient participant engagement and data quality. After exclusions, we retained 425 valid responses for analysis. Table~\ref{tab:demographics} provides an overview of participant demographics as made available by Prolific. Additionally, group-wise balance checks were conducted (ANOVA for age; chi-square tests for sex and ethnicity), confirming no demographic differences across the 3×3 experimental conditions.

The median completion time was 12.1 minutes. Participants received a base payment of \$2 (USD), with an additional performance-based bonus of \$0.10 per correct answer in the task. Given the 30-question task, participants could earn up to \$3 in bonuses, on top of the \$2 base payment. The study was approved by our institution's IRB, and informed consent was obtained from all participants.

\subsection{Task}
\label{Experiment - Task}

    \begin{figure}[htpb]
        \centering
        \includegraphics[width=0.45\textwidth]{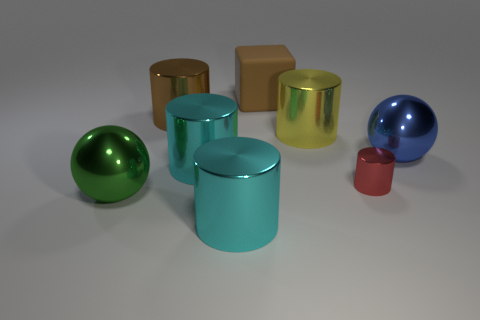}
        \caption{Example CLEVR image used in the experimental task; additional sample questions with answer options are provided in Appendix~\ref{appendix-examples}.}
        \Description{A 3D scene from the CLEVR dataset. Several colored objects of different shapes and materials are arranged on a gray surface against a neutral background, including spheres, cubes, and cylinders in various colors such as blue, red, and green. This scene is used to prompt participants with counting questions in the experiment.}
        \label{fig:clevr_example}
    \end{figure}

Participants completed a visual reasoning task adapted from the CLEVR dataset \cite{johnson_clevr_2017}. The task consisted of counting questions presented over synthetic 3D scenes (e.g., “How many spheres are the same material as the blue cylinder?”), with one question per image and five numeric response options. We restricted prompts to the standard CLEVR attributes - \textit{size} (small/large), \textit{material} (rubber/metal), \textit{shape} (cube/sphere/cylinder), and a constrained \textit{color} set (gray, blue, brown, yellow, red, green, purple) - to minimize linguistic ambiguity and remove the need for domain knowledge. A legend listing these attributes was introduced to participants before the task.

\begin{figure*}[t]
      \centering
      \includegraphics[width=1\textwidth, trim=30 50 30 50, clip]{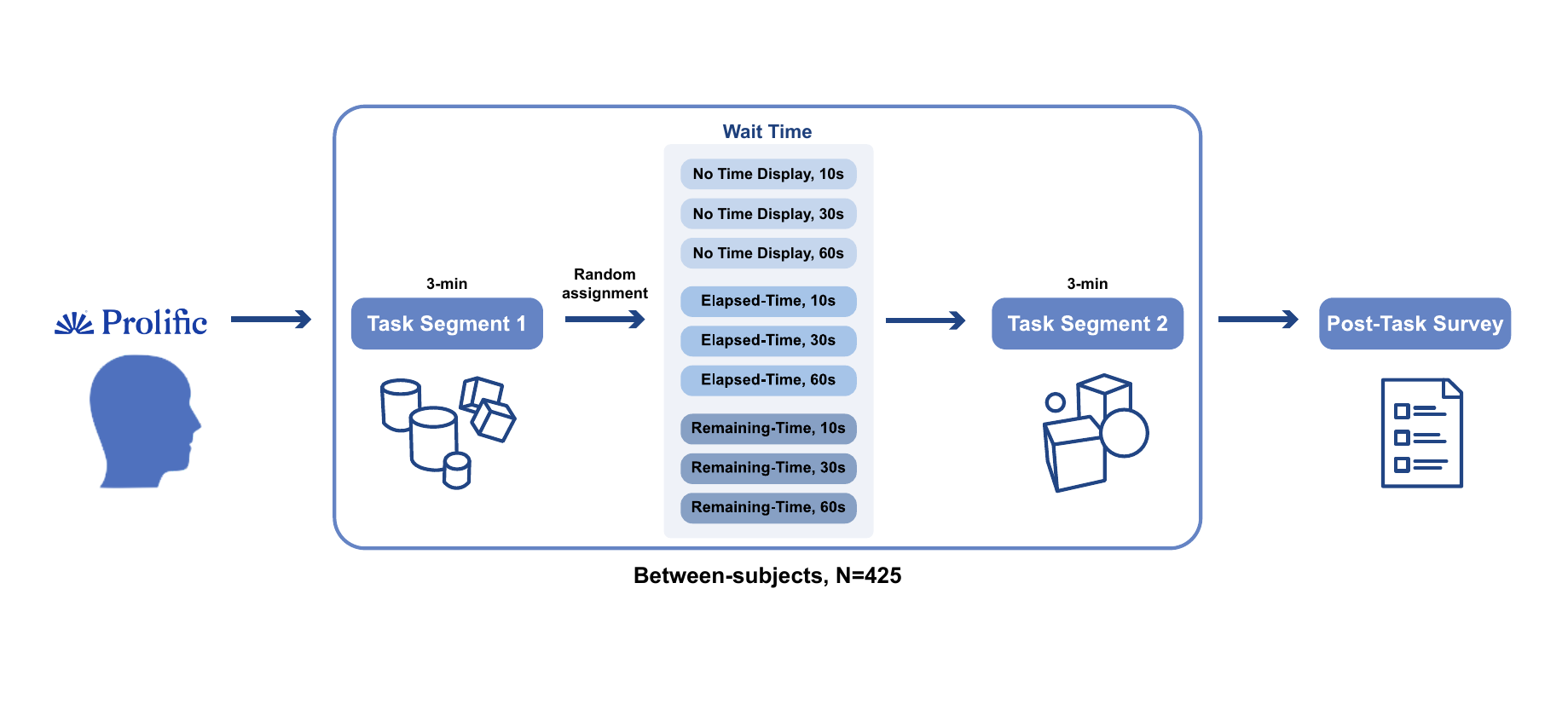}
      \caption{Overview of the study workflow. Participants (N = 425; between-subjects) completed a visual reasoning task in two 3-min segments separated by a system-imposed wait. We crossed \emph{Temporal Feedback Mode} (\emph{No Time Display}, \emph{Elapsed-Time}, \emph{Remaining-Time}) with \emph{Wait Duration} (10 s, 30 s, 60 s) and randomly assigned participants to one of nine conditions. After Segment 2, a post-task survey assessed workload, emotional state, perceived time, and interpretations of the wait.}
      \label{fig:StudyDesign}
      \Description{Overview of the study design (3×3 between-subjects). Two 3-min task segments are separated by a system-imposed wait; N = 425 randomly assigned. The wait crosses feedback mode (\emph{No Time Display}, \emph{Elapsed-Time}, \emph{Remaining-Time}) with duration (10 seconds, 30 seconds, 60 seconds). A post-task survey follows.}
    \end{figure*}

We randomly sampled 30 images (one question per image), split into two 15-question segments separated by the experimental wait (Figure~\ref{fig:StudyDesign}). Each item had five numeric options (0–4) with a single correct answer. Figure~\ref{fig:clevr_example} shows an example visual stimulus presented to participants; additional sample questions with answer options are provided in Appendix~\ref{appendix-examples}.

\paragraph{\textbf{Task Rationale.}}

    We selected a visual reasoning task that relies primarily on perceptual attention and visual working memory while minimizing demands on verbal or semantic processing. This choice aligns with the resource dimensions outlined in Multiple Resource Theory \cite{wickens_multiple_2008}, which differentiates resource pools by perceptual modality (e.g., visual vs.\ auditory) and processing code (e.g., spatial vs.\ verbal). Tasks engaging visual-spatial resources are less confounded by linguistic variability and frequently used to examine attentional load, performance degradation, and latency effects in HCI and cognitive science (e.g., \cite{liu_effects_2014, ballard_memory_1995}). Visual–spatial paradigms have also been common in interruption research across HCI and cognitive psychology (e.g., \cite{altmann_memory_2002, monk_effect_2008}), where they allow precisely timed disruptions and support metrics like resumption lag and performance degradation after delay.

    Within the visual–spatial class, CLEVR-style questions provide structured visual reasoning challenges that draw on selective attention, visual comparison, and short-term memory \cite{johnson_clevr_2017}. While originally designed to benchmark compositional reasoning in machine learning systems, the dataset’s consistent object structure and syntactic simplicity make it well-suited for controlled experimentation. Each question is brief, independently scorable, and grounded in perceptual operations, which supports reliable measurement while reducing interpretive variability and knowledge expertise effects.

\paragraph{\textbf{Scope and Limits of the Task Paradigm.}}
    The generalizability of this task setup is bounded by its simplified structure. Rather than simulating the complexity of real-world workflows such as collaborative editing, game play, or document authoring, the experiment isolates perceptual–attentional mechanisms involved in re-engaging attention. As aforementioned, this mirrors common practice in HCI and psychophysics \cite{cutrell_effects_2000, ratwani_spatial_2008}, where tightly controlled tasks are used to investigate timing, interruption, and attentional effects with internal validity. Although these paradigms not fully capture the richness of real-world workflows, they help isolate perceptual and behavioral patterns that can inform the design of more complex, real-world systems.

\subsection{Procedure}
Participants were recruited via Prolific and completed the study using a desktop or laptop computer. After providing informed consent, participants reviewed instructions about the task. Participants then completed a trial session with three practice questions. Upon submission, they received immediate answer feedback and explanations for each response. To motivate performance in the main session, participants were reminded that:

\begin{itemize}
    \item \textbf{Accuracy Incentive:} Bonus payments of \$0.10 are awarded per correct answer (max \$3.00 across the 30 questions), in addition to a \$2.00 base payment.
    \item \textbf{Speed Incentive:} Each page (comprising 15 questions per segment) had a 3-minute limit to prevent indefinite completion. Pages auto-submitted upon timeout.
\end{itemize}

After the practice trial, participants began the main task, which comprised two 15‐question segments.  Upon completing the first segment (with its questions on a single-page continuous scroll) and selecting “Next,” participants were shown a wait screen. Depending on the assigned condition, this wait varied in duration (\emph{10s}, \emph{30s}, \emph{60s}) and temporal feedback mode (\emph{No Time Display}, \emph{Elapsed‐Time}, \emph{Remaining‐Time}). Once the wait elapsed, the interface automatically advanced participants to Segment 2. This structure ensured that the transition from Segment~1 to Segment~2 unfolded as a continuous task flow, with the wait mimicking a natural page-transition delay.

Upon submission of both task segments, participants completed a post-experiment survey that measured their overall perceptions of the session (Sections \ref{Measure-perception of wait} - \ref{measures-qualitative reflection}), capturing global evaluations of workload, affect, and perceived duration of the wait, rather than segment-specific states. This survey was administered only at the end of the session to maintain task continuity and avoid introducing an additional interruption between segments. Because the two–segment task was experienced as a single continuous activity and the wait was the only experimental manipulation, these end-of-session measures reflect how the temporal feedback and wait shaped participants’ overall experience.

An embedded attention check was used to screen for inattentive responses. The median completion time for the entire study was approximately 12 minutes.

\subsection{Experimental Design}
\label{Section 3.4 - Experimental Design}

Participants were randomly assigned to one of nine experiment cells in a 3\,$\times$\,3 between-subjects design crossing \textbf{Temporal Feedback Mode} with \textbf{Wait Duration}:

\begin{figure*}[t]
  \centering
  \includegraphics[width=0.9\textwidth]{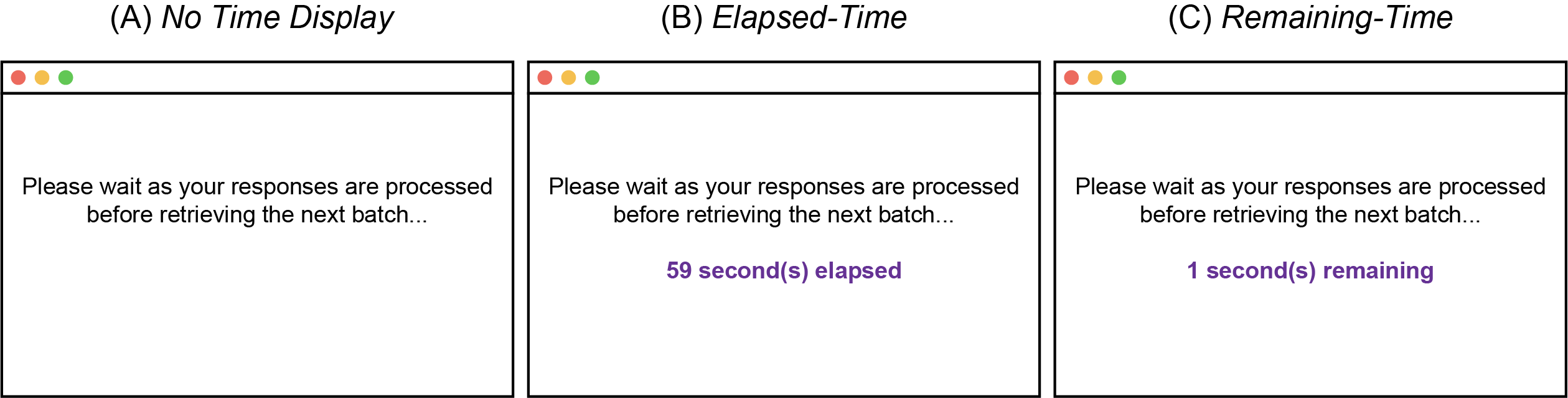}
  \caption{Three experimental wait-screen conditions varying in \emph{Temporal Feedback Mode}: (A) \emph{No Time Display}, (B) \emph{Elapsed-Time}, and (C) \emph{Remaining-Time}. The numerical display (when present) updates once per second (1 Hz), and all conditions show the same base message during the wait. Each feedback mode was paired with one of three wait durations (10s, 30s, 60s).}
  \Description{Temporal feedback modes for wait screens. (A) \emph{No Time Display}: message only. (B) \emph{Elapsed-Time}: seconds elapsed, updating once per second. (C) \emph{Remaining-Time}: seconds remaining, updating once per second. All versions share the same message; waits are 10 seconds, 30 seconds, or 60 seconds by condition.}
  \label{fig:CountType}
\end{figure*}

\subsubsection{\textbf{Temporal Feedback Mode}} 

\begin{enumerate} 
    \item \emph{No Time Display}: No timer is shown; only a neutral wait message appears. Provides zero temporal information; the system neither tracks the passage of time nor reveals the endpoint. 
    \item \emph{Elapsed-Time}: A numeric label shows whole seconds \emph{elapsed} since the wait began (0\,\(\rightarrow\) duration). Provides partial information; the system tracks time passed but does not reveal when the wait will end. 
    \item \emph{Remaining-Time}: A numeric label shows whole seconds \emph{remaining} until the wait ends (duration \(\rightarrow\) 0). Provides complete information; the system tracks time and reveals the endpoint from the outset. 
\end{enumerate}

        Our three temporal-feedback modes differ in the amount of temporal information they provide - none, partial, or complete.  In digital interfaces, this informational level is closely tied to timer \emph{directionality}. A count-up display starts at zero and increments forward; it therefore communicates only \emph{elapsed} time unless an additional cue specifies the total duration. In contrast, a count-down display starts at the total duration and moves toward zero, inherently revealing the \emph{endpoint} from the outset. This pairing reflects how directionality has been implemented in previous work on time perception. For example, Komatsu et al.~\cite{komatsu_waiting_2024} compare count-downs and count-ups using the standard numerical formats in which count-downs make the endpoint visible at onset while count-ups do not provide this information upfront. Their experiments therefore illustrate the same practical co-occurrence of directionality and endpoint visibility that appears in our study design. Because directionality constrains informational level in this way, we describe our conditions in terms of temporal-feedback \emph{modes} (\emph{No Time Display}, \emph{Elapsed-Time}, \emph{Remaining-Time}). Section~\ref{Limitations} elaborates further on how this co-occurrence limits isolating directionality from informational completeness.

\subsubsection{\textbf{Wait Duration}}

(10s, 30s, or 60s)

Illustrated in Figure~\ref{fig:CountType}, the timer (when present) was a plain numeric label that updated once per second (1\,Hz) in whole seconds. There was no progress bar, motion, or other animation. The base message, layout, and styling were identical across conditions; only the numeric string changed each second. When the assigned duration completed, the wait screen dismissed automatically and the second task segment loaded. We used a simple numeric text display so that temporal feedback differed only in the information it conveyed, rather than in visual motion or animation cues known to affect perceived duration independently \cite{gomez_filled-duration_1979, myers_importance_1985}.

        The three wait durations (10s, 30s, 60s) correspond to established latency thresholds in HCI and cognitive psychology. Delays under 10–12s are generally considered tolerable and allow users to maintain task focus without breaking engagement \cite{bailey_need_2006, nielsen_powers_2009}. Around 30s, subjective frustration rises and perceived system responsiveness declines, marking a shift from tolerable latency to disruptive wait \cite{selvidge_how_1999, nah_study_2004}. By 60s, users often disengage or assign negative interpretations to the delay, such as assuming system failure or poor content quality \cite{ramsay_psychological_1998, nielsen_powers_2009}. Selecting 10/30/60s therefore allows us to examine temporal-feedback effects across distinct perceptual thresholds without entering very short (<5\,s) or very long (>90s) waits that involve different user strategies.

    These durations also suit the structure of the implemented visual–spatial task. Because the items are brief, uniform, and independent, pauses of 10–60s allow attention to disengage and later re-engage without potentially altering the cognitive requirements of the task. Participants encountered the wait after completing the first 3-min task segment (comprising 15 items) and received a neutral message (“Please wait as your responses are processed before retrieving the next batch…”), meant to reflect a plausible short system-imposed wait common in screen-based experiences. Longer waits typical of compute-intensive workflows (e.g., large-scale compilation, model training, or data processing) fall outside the scope of this study; understanding temporal feedback in those contexts would be a valuable extension. These longer waits often encourage multitasking or activity switching rather than within-task resumption \cite{nah_study_2004}. We discuss how expectations about wait duration vary across different real-world tasks, and how these expectations may affect subjective experience in Section~\ref{Limitations}.

\subsection{Experimental Platform} 
\label{Experimental Platform}

The experiment was hosted on Qualtrics\footnote{\url{https://www.qualtrics.com}}; participants were directed to the Qualtrics survey after accepting the task on Prolific, and were automatically redirected back to Prolific upon completion. All task instructions, visual reasoning items, wait-time manipulations, and post-task surveys were implemented within Qualtrics using built-in logic and custom JavaScript. The system-imposed wait periods were programmed using JavaScript timers that dynamically displayed the feedback text (e.g., elapsed or remaining time) based on the assigned condition. All experimental data, including condition assignment, task and survey responses, were recorded within Qualtrics for analysis.

\subsection{Measures}

\subsubsection{Task Performance}

Task performance was defined as the percentage of correctly answered items within each task segment. Each segment contained 15 short, objective, multiple-choice items sampled from the CLEVR dataset (Section \ref{Experiment - Task}). Performance per segment was computed as:
\[
\text{Performance (\%)}=\left(\frac{\text{Correct Answers}}{15}\right)\times 100
\]

Unanswered items were counted as incorrect. By indexing performance against the full set of questions, this measure incorporates both correctness and task completion, offering a balanced indicator of speed and accuracy. We excluded participants who answered fewer than 80\% of all questions across task segments (i.e., less than 24 of 30 questions), ensuring the dataset reflected sufficient engagement.

\subsubsection{Perception of Wait}
\label{Measure-perception of wait}

To examine how participants retrospectively experienced and interpreted the system-imposed wait, we included the following questions:

\begin{enumerate}
    \item \textbf{Recall of Wait Time (Yes/No):} “Was there a wait time between the two parts of your task?”
    \item \textbf{Perceived Duration (7-point scale; 1: Very Short, 7: Very Long):} “How long did the wait time feel?” 

\end{enumerate}

Participants who indicated they recalled a wait period (i.e., answered “Yes” to the first question) were shown the remaining Perceived Duration item, as well as the open-ended question (Section \ref{measures-qualitative reflection}). This allowed us to collect reflections on wait perception without assuming conscious awareness of the manipulation.

These items assess how noticeable or cognitively salient the wait was across different feedback modes and durations. Rather than capturing how the wait felt moment-to-moment, they reflect users’ retrospective memory of the delay. Prior work shows that such memories can diverge from real-time experience. For example, faster countdowns may feel shorter during the wait but be recalled as longer afterward \cite{ghafurian_countdown_2020}. Our aim was therefore not to estimate psychophysical thresholds for “short” versus “long” waits, or use these items for addressing core research questions or primary claims. Instead, they provide additional context on how participants registered and remembered the delay.

\subsubsection{Psychological Experience}

To understand how the wait shaped participants’ end-of-session emotional and cognitive evaluations, we administered two validated instruments. These measures were administered once at the end of the full task sequence (Segment 1 → wait → Segment 2) and therefore reflect participants’ overall evaluations of the session rather than segment-specific states. As a result, differences observed in these subjective measures should be interpreted as reflecting how the waiting experience shaped participants’ overall impressions of the session, rather than their moment-to-moment state during the second task segment.

\paragraph{Cognitive Workload:}  

We included the NASA Task Load Index \cite{hart_development_1988}, a widely used instrument to measure perceived workload in task environments. Participants rated the following dimensions on 0–100 sliders:
\begin{itemize}
    \item \textbf{Mental Demand:} How mentally demanding were the tasks? 
    \item \textbf{Physical Demand:} How physically demanding were the tasks? 
    \item \textbf{Temporal Demand:} How hurried or rushed was the pace of the tasks? 
    \item \textbf{Performance:} How successful were you in accomplishing what you were asked to do? 
    \item \textbf{Effort:} How hard did you have to work to accomplish your level of performance? 
    \item \textbf{Frustration:} How insecure, discouraged, irritated, stressed, and annoyed were you?
\end{itemize}

\paragraph{Emotional State:}
We adapted the Pleasure-arousal-dominance (PAD) model \cite{mehrabian_pleasure-arousal-dominance_1996} to assess participants’ affective responses across valence (pleasure), arousal, and dominance (control) dimensions. Prior work has shown PAD to be effective in evaluating user experiences in time-sensitive digital environments \cite{wang_effect_2021}. All items were rated on a 7-point Likert-scale:

\begin{itemize} 
\item \textbf{Pleasantness:} “How pleasant vs. unpleasant was your overall experience?”
\item \textbf{Excitedness:} “How calm vs. excited did you feel while completing the tasks?”
\item \textbf{Sense of Control:} “To what extent did you feel in control vs. out of control?”
\end{itemize}

        \begin{table}[htbp]
            \caption{Overview of study measures}
            \Description{Study measures mapped to research questions. RQ1 - Task Performance: Percent correct in Task Segment 2. RQ2 - Wait Time Perception: recall of wait (Yes/No) and perceived duration (7-point). RQ2 - Cognitive Workload: NASA-TLX dimensions-mental, physical, temporal demand, effort, frustration, performance (0–100 sliders). RQ2 - Emotional State: PAD-pleasantness, excitedness, sense of control (7-point). RQ3 - Free Response: open-ended prompt on how the wait affected experience and next-segment performance.}
            \label{tab:measures}
            \small
            \begin{tabular}{p{2.5cm} | p{5cm}}
            \toprule
            \textbf{Dimension} & \textbf{Measure} \\ \midrule
            Task Performance \newline\textit{(RQ1)} & Score (\%) for Task Segment 2 \\ \midrule
            Wait Time Perception \textit{(RQ2)} & Recall of Wait (Yes/No); Perceived Duration (7-point Likert scale) \\ \midrule
            Cognitive Workload \newline\textit{(RQ2)} & Mental Demand, Physical Demand, Temporal Demand, Effort, Frustration, Performance (0–100 sliders) \\ \midrule
            Emotional State \newline\textit{(RQ2)} & Pleasantness, Excitedness, Sense of Control (7-point Likert scale) \\ \midrule
            Free Response \newline\textit{(RQ3)} & “How do you think the wait time influenced your experience (e.g., emotions, attention, motivation) or your performance in the next part of the task?” \\ \bottomrule
            \end{tabular}
        \end{table}

\subsubsection{Qualitative Reflection}
\label{measures-qualitative reflection}

In addition to structured survey items, we included an open-ended question inviting participants to describe how the wait influenced their task experience: \textit{“How do you think the wait time influenced your experience (e.g., emotions, attention, motivation) or your performance in the next part of the task?”}  

This item was only shown to participants who indicated they recalled a wait. Responses to this question were later analyzed qualitatively to identify recurring patterns in how participants experienced the system-imposed delay.

\section{Results} 
\label{Results}

        \begin{figure*}[htbp]
          \centering
          \includegraphics[width=0.83\linewidth, trim=3pt 15pt 15pt 15pt, clip]{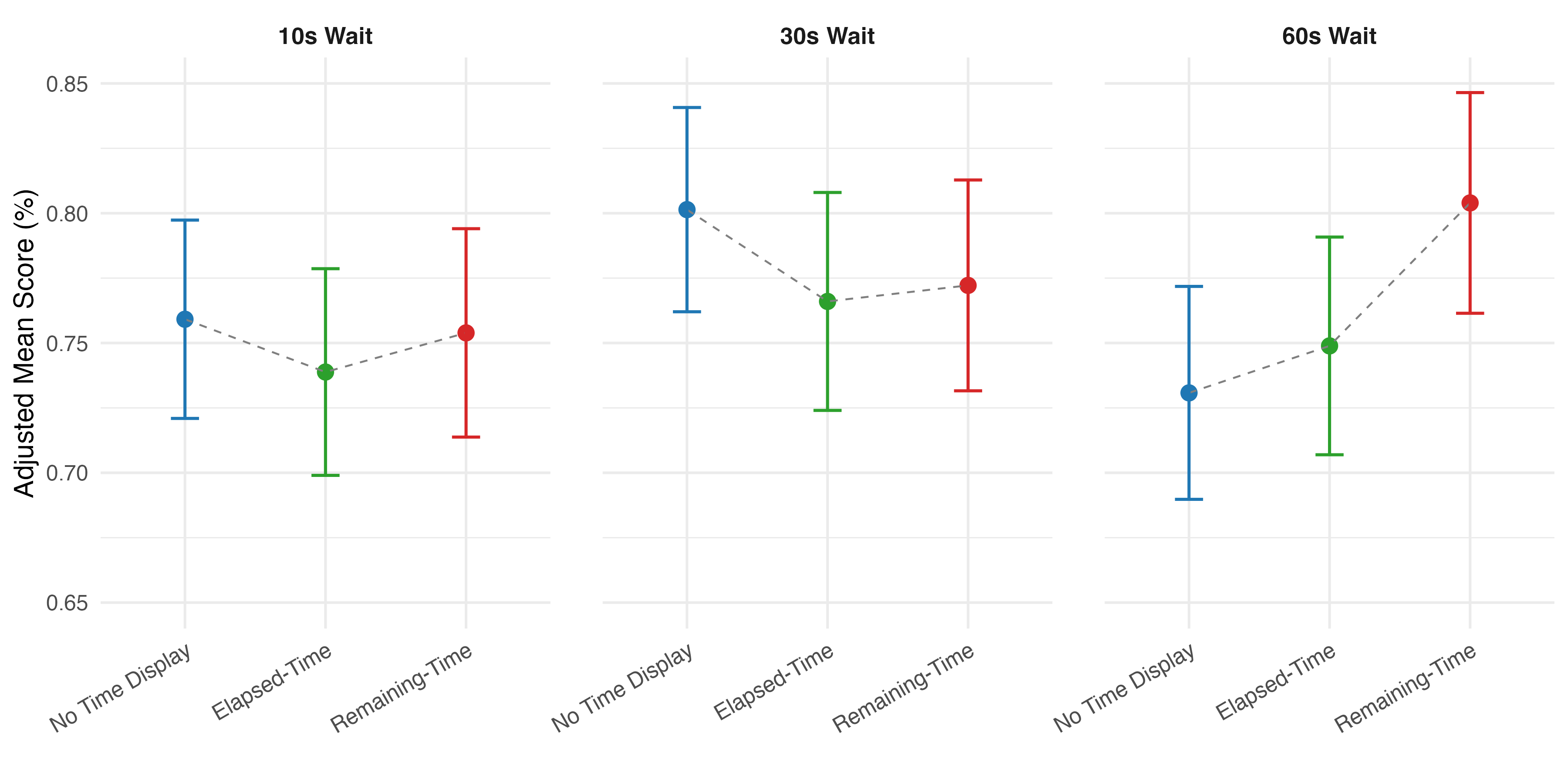}
          \caption{Covariate-adjusted mean performance across Temporal Feedback Modes and Wait Durations from the ANCOVA model. Error bars represent 95\% confidence intervals.}
          \label{fig:performance_interaction}
          \Description{
            The figure contains three panels showing covariate-adjusted mean performance for the three Temporal Feedback Modes at 10\,s, 30\,s, and 60\,s wait durations. 
            Each panel displays one point per feedback mode with 95\% confidence intervals and a dashed line connecting the points. 
            At 10\,s, the adjusted means are similar across modes. 
            At 30\,s, No Time Display is numerically highest. 
            At 60\,s, Remaining-Time is numerically highest. 
            Confidence intervals overlap in all panels.}
        \end{figure*}

For Task Performance, where baseline differences required a covariate-adjusted analysis, we used ANCOVA with Segment~1 as a covariate to evaluate the effects of \textbf{Temporal Feedback Mode} and \textbf{Wait Duration} on post-wait task performance (in Segment~2).

For other quantitative outcomes (including perceived duration, cognitive workload, and emotional state ratings), we conducted between-subjects analyses using aligned rank transform (ART) ANOVA \cite{wobbrock_aligned_2011}. ART ANOVA was selected as a non-parametric alternative appropriate for our 3×3 factorial design, given that many dependent variables either violated normality assumptions (as assessed via Shapiro–Wilk tests) or were measured on ordinal Likert-type scales. Where significant main effects or interactions were observed, we conducted post hoc comparisons using ART contrasts with Tukey-adjusted pairwise comparisons via the \texttt{emmeans} package \cite{lenth_emmeans_2025}. Recall of Wait Time was the only categorical measure analyzed using a chi-squared test of independence.

\subsection{Task Performance}

Our primary performance outcome focused on \textit{Segment~2}, which followed the wait-time manipulation and therefore reflects the downstream impact of Temporal Feedback Mode and Wait Duration on task performance. Before analyzing post-wait performance, we first evaluated baseline equivalence across conditions using Segment~1 performance.

\paragraph{\textbf{Baseline Equivalence Check}} The ART ANOVA on Segment~1 performance showed no significant main effect of Temporal Feedback Mode, $F(2, 416) = 1.63$, $p = .197$, but revealed a significant main effect of Wait Duration, $F(2, 416) = 3.74$, $p = .025$, and a significant interaction, $F(4, 416) = 2.90$, $p = .022$.

As Segment~1 scores significantly differed across conditions, we incorporated Segment~1 scores as a covariate in subsequent analyses to adjust for baseline variability and isolate the effects of the manipulation.

\paragraph{\textbf{ANCOVA with Baseline Adjustment}} Before running the ANCOVA, we verified model assumptions, including normality of residuals and homogeneity of regression slopes. The relationship between Segment~1 and Segment~2 performance varied across wait durations (Segment~1 $\times$ Wait Duration interaction: $F(2, 407)=5.23$, $p=.006$), indicating some slope heterogeneity. However, because Segment~1 was measured prior to random assignment of the wait manipulation, any baseline differences necessarily arose by chance rather than reflecting systematic group differences. Under these conditions, ANCOVA remains appropriate and is widely regarded as a conservative adjustment method \cite{miller_misunderstanding_2001}.

Controlling for Segment~1 performance, the ANCOVA revealed no significant main effects of Temporal Feedback Mode or Wait Duration, nor a significant interaction between the two factors (all $p>.05$). As expected, Segment~1 performance strongly predicted Segment~2 performance, $F(1, 415)=295.11$, $p<.001$, indicating stability in individual performance across segments.

Although the interaction was not statistically significant, the adjusted means provide descriptive contrasts across conditions. Most notably, performance was higher for \emph{Remaining-Time} during longer 60s waits, whereas during shorter waits of 10s and 30s, the \emph{No Time Display} condition did not show reduced performance relative to the other modes (see Figure~\ref{fig:performance_interaction}).

\subsection{Perception of Wait}

       \begin{figure}[b]
           \centering
         
           \begin{subfigure}[t]{0.48\textwidth}
             \centering
             \raisebox{11pt}{\includegraphics[width=\linewidth]{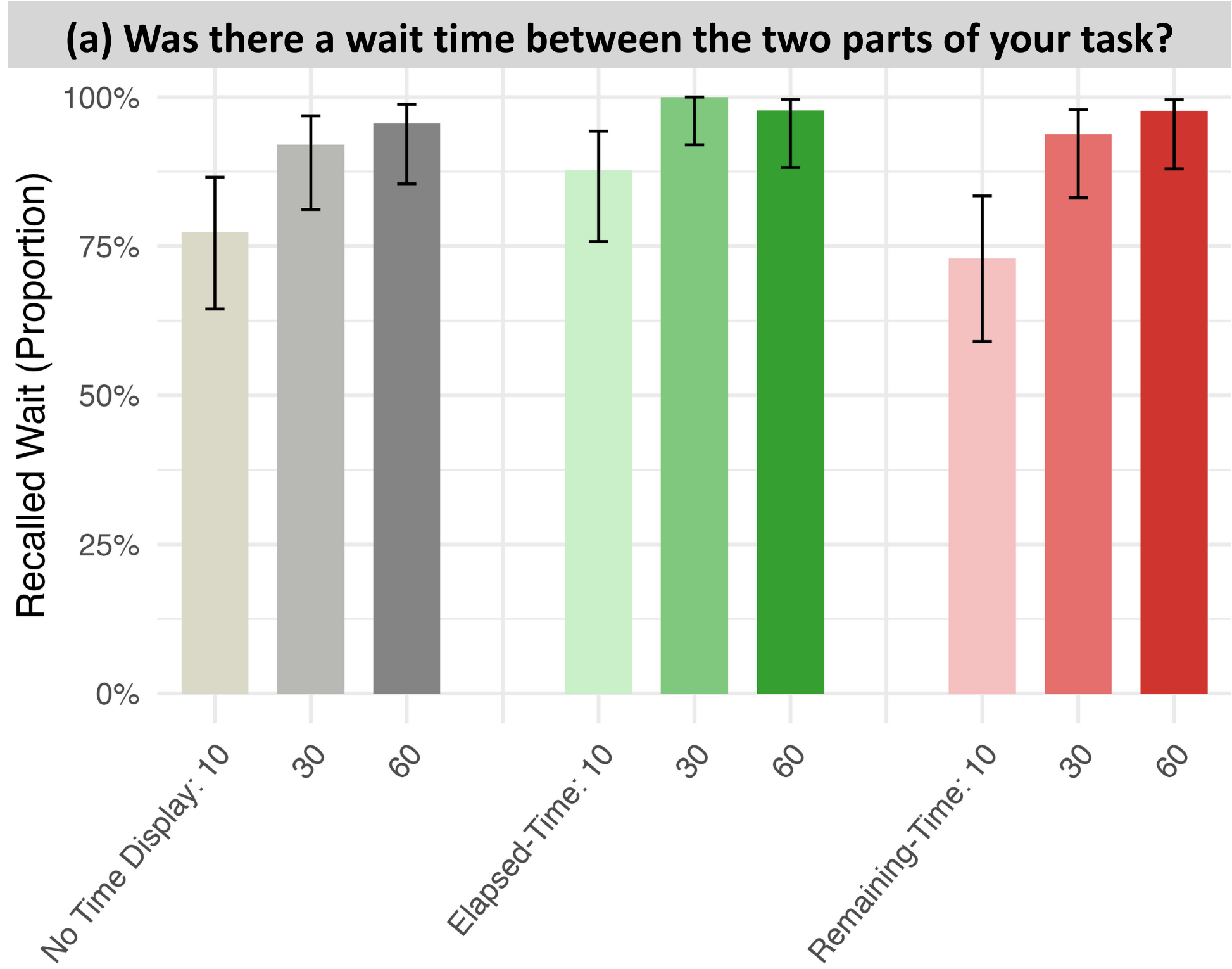}} 
             \caption*{}  
           \end{subfigure}
           \hfill
           
           \begin{subfigure}[t]{0.5\textwidth}
             \centering
             \includegraphics[width=\linewidth]{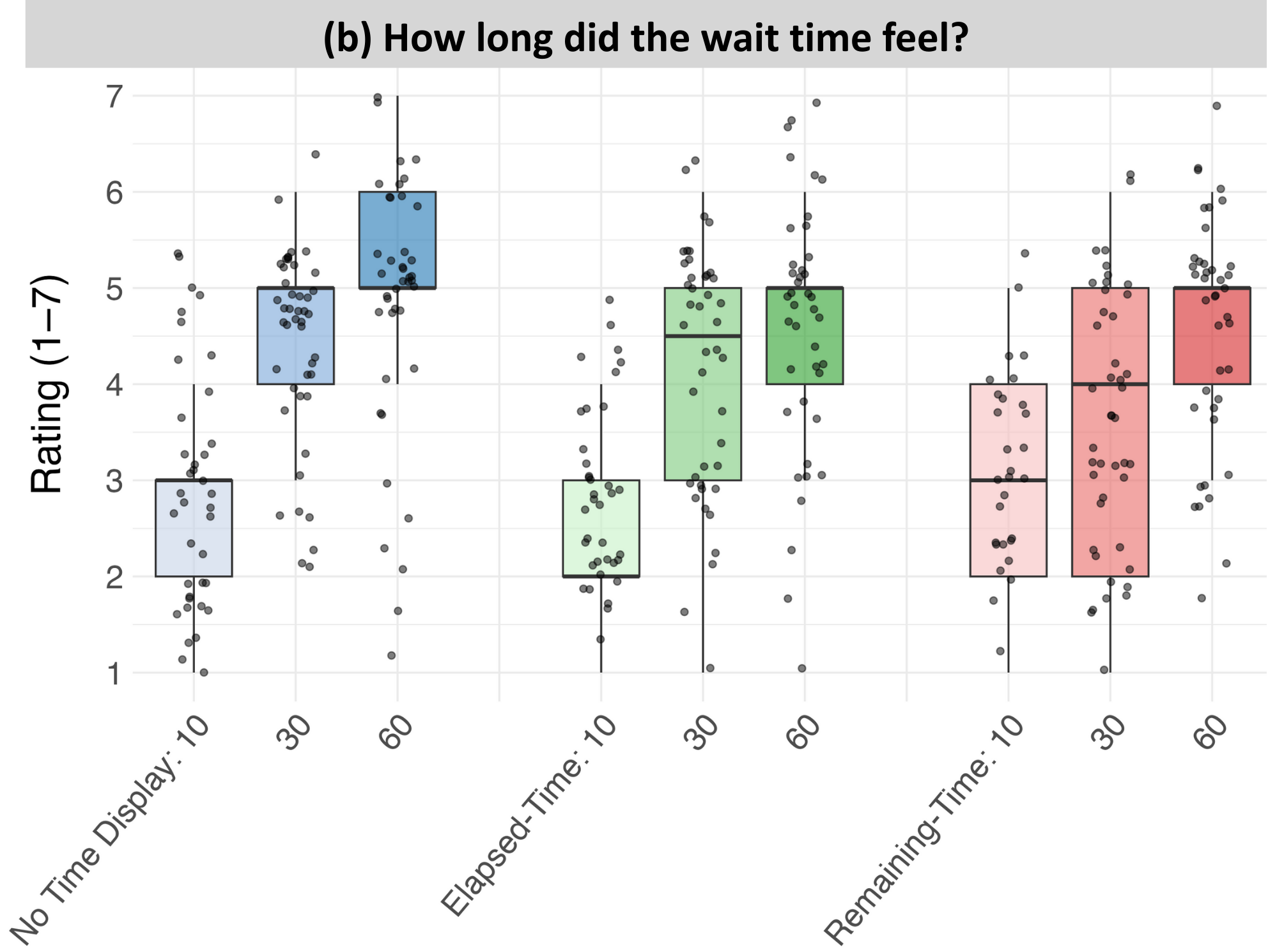}
             \caption*{} 
           \end{subfigure}
     
          \caption{Summary of wait-perceptual outcomes: (a) Proportion of participants who recalled a wait, with error bars representing 95\% confidence intervals (Wilson method). (b) Perceived duration ratings on a 7-point scale.}
          \label{fig:perception_plots}
          \Description{Wait perception outcomes. (a) Recall increases with longer waits: lowest at 10 seconds and near-ceiling at 30–60 seconds across \emph{No Time Display}, \emph{Elapsed-Time}, and \emph{Remaining-Time}; 10 seconds is lowest under \emph{No Time Display}. (b) Perceived duration rises monotonically from 10 seconds to 60 seconds for all feedback modes; differences between modes are small relative to duration. Bars in (a) show proportions with 95\% Wilson Confidence Intervals; (b) shows boxplots (median line, IQR box, points = participants).}
        \end{figure}

\paragraph{Recall of Wait.}
Participants' ability to recall the wait period was strongly associated with condition ($\chi^2(8, N=425) = 38.9$, $p < .001$, Cramér’s $V = .30$). Recall increased with longer durations (Figure~\ref{fig:perception_plots}), reaching above 95\% in all \emph{60s} conditions, while at \emph{10s}, recall was lower: 77.4\% in \emph{No Time Display}, 87.8\% in \emph{Elapsed‐Time}, and 72.9\% in \emph{Remaining‐Time}. This confirms that longer waits are more salient and more likely to be remembered.

\paragraph{Perceived Duration.}
Perceived wait duration was influenced by both feedback mode and wait length. An ART ANOVA revealed a significant main effect of \textbf{Temporal Feedback Mode} ($F(2, 374) = 5.21$, $p = .006$, $\eta^2_p = .027$) and \textbf{Wait Duration} ($F(2, 374) = 85.21$, $p < .001$, $\eta^2_p = .314$), but no significant interaction (Figure \ref{fig:perception_plots}). As expected, longer waits were perceived as longer (60s: $M = 4.74$, $SD = 1.37$, CI[4.50, 4.97]; 30s: $M = 3.99$, $SD = 1.29$, CI[3.77, 4.21]; 10s: $M = 2.69$, $SD = 1.19$, CI[2.47, 2.91]; all $p < .001$), confirming that subjective ratings tracked objective durations. 
Among feedback types, \emph{No Time Display} ($M = 4.06$, $SD = 1.54$, CI[3.79, 4.33]) was perceived as significantly longer than \emph{Remaining‐Time} ($M = 3.70$, $SD = 1.48$, CI[3.43, 3.96]; $p < .01$), suggesting that lacking a visible timer made waits feel longer. There were no other pairwise differences.

\subsection{Psychological Experience}

\subsubsection*{\textbf{Cognitive Workload (NASA-TLX)}}

\begin{figure*}[htbp] 
            \centering 
            \includegraphics[width=1\textwidth]{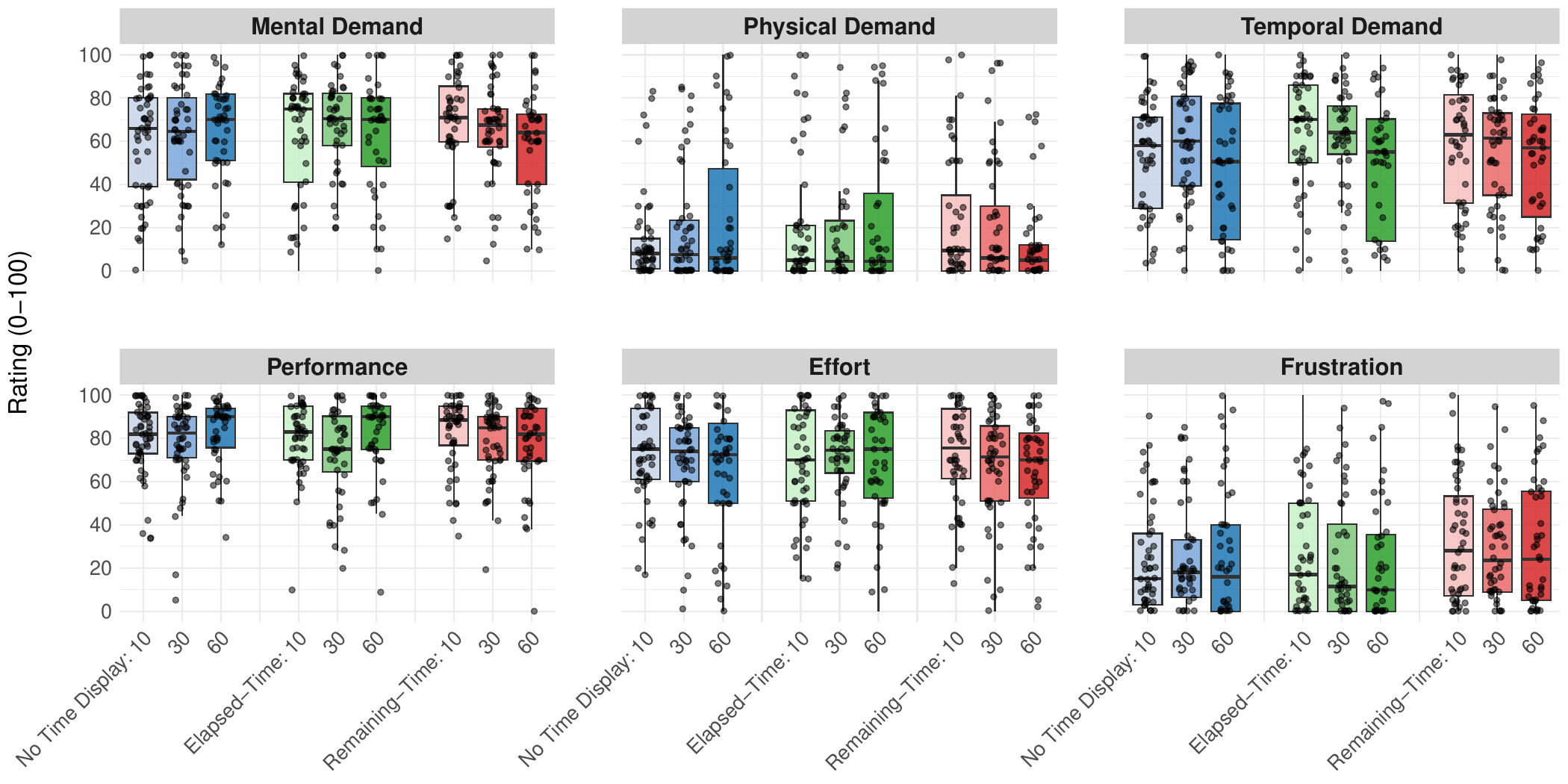} 
            \caption{NASA–TLX workload ratings across all conditions (3 \textbf{Temporal Feedback Modes} × 3 \textbf{Wait Durations}). Each panel shows one dimension - Mental Demand, Physical Demand, Temporal Demand, Performance, Effort, Frustration - using boxplots with jittered participant points (0–100 scale).} 
            \label{fig:nasa_tlx_boxplot} 
            \Description{Boxplots displaying NASA-TLX workload ratings across nine experimental conditions, organized by six dimensions: Mental Demand, Physical Demand, Temporal Demand, Performance, Effort, and Frustration.} 
        \end{figure*}

\paragraph{Temporal Demand.}  
There was a significant main effect of \textbf{Wait Duration} on temporal demand, $F(2, 416) = 4.34$, $p < .05$, $\eta^2 = .020$. No main effect of \textbf{Temporal Feedback Mode} or interaction effect was observed. Participants in the \emph{60s} condition reported significantly lower perceived time pressure ($M = 48.7$, $SD = 31.5$, CI[43.3, 54.1]) compared to \emph{30s} ($M = 58.7$, $SD = 25.9$, CI[54.4, 63.0]; $p = .025$) and \emph{10s} ($M = 57.7$, $SD = 28.3$, CI[53.1, 62.3]; $p = .032$).

\paragraph{Frustration.}  
A significant main effect of \textbf{Temporal Feedback Mode} was found for frustration, $F(2, 416) = 4.20$, $p < .05$, $\eta^2 = .020$. \emph{Remaining‐Time} ($M = 30.7$, $SD = 27.4$, CI[26.1, 35.3]) elicited significantly higher frustration than \emph{Elapsed‐Time} ($M = 24.1$, $SD = 27.9$, CI[19.4, 28.8]; $p = .019$).

\paragraph{Performance (perceived).}  
A borderline effect of \textbf{Wait Duration} was observed on self-rated performance, $F(2, 416) = 2.99$, $p = .051$, $\eta^2 = .014$. Participants rated their performance higher in the \emph{60s} ($M = 81.2$, $SD = 18.7$, CI[78.0, 84.4]) condition than the \emph{30s} condition ($M = 76.7$, $SD = 19.6$, CI[73.5, 80.0]; $p = .049$).

\paragraph{Mental Demand, Physical Demand, and Effort.}  
No significant effects (p > .05) were found for these workload dimensions.

\subsubsection*{\textbf{Emotional State}}
          
    \begin{figure*}[t] 
        \centering 
        \includegraphics[width=1\textwidth]{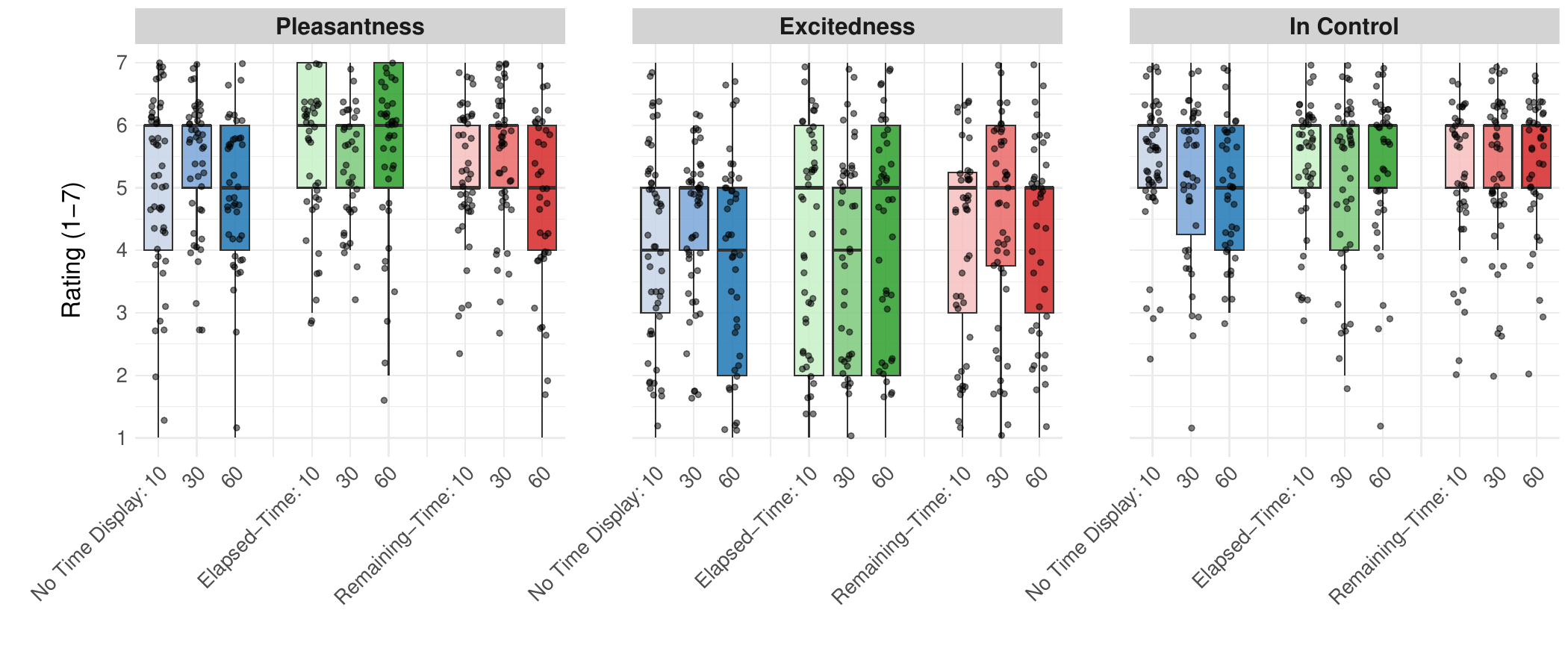} 
        \caption{PAD emotional-state ratings across conditions (1–7 scale). Panels show (A) Pleasantness, (B) Excitedness, and (C) Sense of Control across the 3×3 combinations of \textbf{Temporal Feedback Mode} (\emph{No Time Display}, \emph{Elapsed-Time}, \emph{Remaining-Time}) and \textbf{Wait Duration} (10s, 30s, 60s). Boxplots with jittered participant points.} 
        \label{fig:pad_boxplot} 
        \Description{PAD ratings by condition (1–7). Panels show (A) Pleasantness, (B) Excitedness, and (C) Sense of Control across the 3×3 combinations of Temporal Feedback Mode and Wait Duration. Each panel contains nine boxplots (3 Temporal Feedback Modes × 3 Wait Durations) with jittered participant points.}
    \end{figure*}

\paragraph{Pleasantness.}  
A main effect of \textbf{Temporal Feedback Mode} was found for pleasantness ratings, $F(2, 416) = 3.51$, $p < .05$, $\eta^2_p = .017$.
Post hoc comparisons revealed that participants in the \emph{Remaining‐Time} condition ($M = 5.33$, $SD = 1.29$, CI[5.11, 5.55]) rated the experience as significantly less pleasant than those in the \emph{Elapsed‐Time} condition ($M = 5.67$, $SD = 1.18$, CI[5.47, 5.87]; $p = .025$). No other pairwise differences reached significance, though descriptively, \emph{No Time Display} ($M = 5.33$, $SD = 1.29$, CI[5.12, 5.54]) was rated similarly low in pleasantness as \emph{Remaining‐Time}.

This aligns with the frustration results, where \emph{Elapsed‐Time} also elicited significantly lower frustration than \emph{Remaining‐Time}, reinforcing that lower frustration was associated with higher pleasantness in these conditions.

\paragraph{Excitedness, and Sense of Control.}  
No significant effects (p > .05) were found for these two dimensions.

\subsection{Perceived Influence of the Wait (Qualitative Reflection)}
\label{Results - Qual}

    \begin{table*}[t]
        \centering
        \caption{Seven themes from participants’ open-ended responses (n = 383). Counts reflect responses coded with each theme; multiple codes per response were allowed, so percentages do not sum to 100\%.}
        \Description{Themes from open-ended responses (n = 383; multiple codes allowed). Most common were \emph{No Influence} 132 (34.5\%) and \emph{Task Re-Engagement} 97 (25.3\%), followed by \emph{Restorative Pause} 64 (16.7\%). Less frequent were \emph{Affective Dysregulation} 57 (14.9\%), \emph{Affective Regulation} 41 (10.7\%), and two smaller themes-\emph{Cognitive Disruption} 21 (5.5\%) and \emph{Interpretive Ambiguity} 21 (5.5\%).}
        \label{tab:wait_themes}
        \begin{tabular}{p{3cm} p{5.4cm} p{6.6cm} p{1.4cm}}
            \toprule
            \textbf{Theme} & \textbf{Brief Description} & \textbf{Example Quote} & \textbf{Count (\%)} \\
            \midrule
            No Influence & The wait had no noticeable effect on experience or performance. & “I didn't feel any effect.” & 132 (34.5\%) \\[4pt]
            Restorative Pause & The wait acted as a mental break or recovery period. & “It gave me a quick second to relax before the next round.” & 64 (16.7\%) \\[4pt]
            Task Re‐Engagement & The wait supported motivation or planning for the next segment. & “It made me feel more motivated to do well.” & 97 (25.3\%) \\[4pt]
            Affective Regulation & The wait helped users calm down or reduce stress. & “Helped me calm down and gather myself.” & 41 (10.7\%) \\[4pt]
            Affective Dysregulation & The wait generated negative arousal (stress, anxiety, impatience). & “Made me more anxious.” & 57 (14.9\%) \\[4pt]
            Cognitive Disruption & The wait interrupted users’ flow or focus. & “I was ‘in the zone’ and the wait pulled me out of it.” & 21 (5.5\%) \\[4pt]
            Interpretive Ambiguity & The wait caused confusion about system state (“Is it still working?”). & “I thought there might have been an issue with the study at first.” & 21 (5.5\%) \\
            \bottomrule
        \end{tabular}
    \end{table*}

    We conducted a thematic analysis \cite{braun_using_2006} of participants’ open-ended responses to the question: \emph{“How do you think the wait time influenced your experience (e.g., emotions, attention, motivation) or your performance in the next part of the task?”} The first author inductively open-coded a stratified seed set to draft a seven-theme codebook (Appendix~\ref{appendix-qualitative codebook}). Two trained  coders (non-authors) then independently applied the finalized codebook to the full set of free-text responses (binary present/absent per theme; n = 383). During coding, ambiguities were discussed and resolved to refine definitions, and updates were back-applied to ensure uniform use of the codebook. Inter-rater reliability was computed on the two coders' independent labels under the finalized definitions; Cohen’s kappa indicated substantial agreement per theme ($\kappa = 0.69$-$0.85$).
    
    Table~\ref{tab:wait_themes} summarizes the seven themes, including example quotes and overall distribution of theme frequencies. Some responses were assigned to multiple themes; thus, percentages do not sum to 100\%.

\begin{description}
    \item[No Influence]
    was the most frequent theme overall. It was especially common in \emph{Elapsed-Time} (38.5\%) and \emph{No Time Display} (37.4\%) feedback modes, suggesting that when time feedback was either absent or presented only as elapsed time, waits often felt inconsequential. As expected, "No Influence" responses were most prevalent in short waits (39.5\% at 10s), and decreased at 30s (31.9\%) and 60s (32.6\%).
    
    \item[Task Re-Engagement]
    was the second most common theme. It was most prominent in the \emph{Remaining-Time} feedback mode (27.9\%), followed by \emph{Elapsed-Time} (26.9\%) and \emph{No Time Display} (21.4\%), suggesting that the absence of time feedback was least likely to prepare participants for the upcoming task segment. The Task Re-Engagement theme was relatively similar at 10s (27.7\%) and 30s (27.4\%), then lowering at 60s (20.9\%). This suggests that the potential for waits to facilitate task re-engagement diminishes during longer delays.

    \item[Restorative Pause.]
    Themes were relatively stable across wait durations but were more frequent in the \emph{Remaining-Time} condition, suggesting that countdown feedback may help participants reframe the wait as a useful or restorative break regardless of duration.
    
   \item[Affective Dysregulation]
    peaked in the \emph{No Time Display at 30s} condition, suggesting that negative emotions such as frustration and impatience were most likely when waits were moderate and no temporal feedback was provided. Interestingly, at 60 seconds, dysregulation eased, possibly because participants had more time to emotionally adjust or disengage, whereas 30 seconds was long enough to provoke frustration but too short to fully adapt.
    
    \item[Affective Regulation.]
    Themes of affective regulation (e.g., calming or stress reduction) were mentioned least in \emph{No Time Display} conditions, particularly at 10 seconds and 60 seconds, suggesting that the uncertainty of time made positive emotional coping less likely. In contrast, it was most frequent in the \emph{Remaining-Time at 30s} condition, indicating that moderate waits paired with predictable temporal feedback (via a countdown timer) might offer participants an opportunity to actively regulate emotions.

    \item[Cognitive Disruption]
    increased steadily with wait duration, from 1.7\% at 10 seconds to 6.7\% at 30 seconds and 7.8\% at 60 seconds, reflecting greater difficulty maintaining focus after longer interruptions. Across feedback modes, its frequency was relatively similar (5.7-6.2\%), indicating that duration played a stronger role than feedback type in this disruption.
    
    \item[Interpretive Ambiguity]
    was most frequent in longer \emph{No Time Display} conditions, reflecting confusion about the system state (e.g., "Is the webpage still working?").

\end{description}

These qualitative patterns provide rich context to understand participants’ subjective experiences of waiting. In the following Discussion section, we integrate these insights with our quantitative findings to highlight broader implications for design and theory.

\section{Discussion}
\label{Discussion}

Our study examined how different Temporal Feedback Modes shape users’ experience of varying Wait Durations and their ability to resume an ongoing task. Across 425 participants, Feedback Mode and Wait Duration significantly shifted perceived duration, frustration, pleasantness, and temporal demand. However, these experiential differences did not translate into improved post-wait task performance.

We organize the discussion around four insights: (i) the interaction between Temporal Feedback Mode and Wait Duration; (ii) how Feedback Mode modulates perceived duration, time pressure, and uncertainty; (iii) why \emph{Remaining-Time} feedback heightens negative affect more than \emph{Elapsed-Time}; and (iv) why these experiential differences do not translate into differences in post-wait task performance. We then reflect on generalizability and implications for temporal design in interactive systems.

\subsection{\textbf{Insight 1:} No Interaction Between Temporal Feedback Mode and Wait Duration Across Outcomes}

Although both Temporal Feedback Mode and Wait Duration shaped different aspects of subjective experience, their effects appeared to be independent. In other words, we found no evidence that Feedback Mode altered the impact of duration, or that duration modified the influence of Feedback Mode. This directly addresses RQ1 by showing no evidence of a Temporal Feedback Mode $\times$ Wait Duration interaction on post-wait task performance, and it informs RQ2 by showing the same lack of interaction across the subjective experience measures.

This pattern is theoretically informative. Prior HCI and timing research shows that temporal cues shape affective and attentional responses during waits \cite{reed_use_2004, gomez_filled-duration_1979} and that visual feedback can alter perceived waiting and emotional appraisal \cite{lallemand_enhancing_2012, fang_effects_2022}. Our findings refine this view by showing that, within single, deterministic delays embedded in a repetitive and low-complexity visual reasoning task, Feedback Mode and Wait Duration appear to engage separable evaluative processes. Duration effects align with models of temporal estimation driven by attention to the passage of time \cite{treisman_temporal_1963, zakay_role_1996}, whereas mode effects reflect emotional appraisal and the attentional framing provided by interface cues. Empirical work on progress indicators supports this distinction: temporal cues can help maintain focus or prepare for task resumption \cite{liu_supporting_2014, chen_effects_2025} without necessarily interacting with the length of the delay. In our data, these processes produced additive rather than interactive shifts in subjective outcomes.

From a design standpoint, this suggests that, at least within short visual reasoning workflows, Feedback Mode and Wait Duration may be treated as separable levers. Adjusting one does not appear to fundamentally change the influence of the other in this experimental context. This distinction is important for interpreting subsequent findings, which reflect main effects of duration and main effects of Feedback Mode rather than duration-by-mode trade-offs.

\subsection{\textbf{Insight 2:} Temporal feedback modulates perceived duration, time pressure, and uncertainty}
            
Perceived duration and time pressure showed systematic patterns that were not mirrored in task performance. Participants reliably judged longer waits as longer, and \emph{No Time Display} made waits feel significantly longer than \emph{Remaining-Time}. These results are consistent with classic findings that the absence of temporal cues inflates subjective duration because uncertainty increases cognitive load and draws attention toward time itself \cite{zakay_role_1996, gomez_filled-duration_1979}. The qualitative accounts align with this interpretation. Experiences coded as \emph{Interpretive Ambiguity}, such as “I thought there might have been an issue with the study at first,” were most common in \emph{No Time Display} at longer waits, suggesting that uncertainty about system progress contributed directly to perceived duration inflation.

Perceived time pressure (Temporal Demand from NASA-TLX) followed a different pattern. Participants reported the lowest temporal demand at 60 seconds, followed by 30 seconds, then 10 seconds. This pattern suggests that when waits are embedded within a short, repetitive workflow, longer pauses may diffuse a sense of urgency by providing enough time to disengage and reorient, whereas shorter waits sustain a sense of pacing and time pressure. This interpretation is supported by the \emph{Restorative Pause} and \emph{Affective Regulation} themes, which were more common at 60 seconds.

Together, these findings advance RQ2 by showing that temporal feedback and Wait Duration primarily shifted retrospective duration judgments and perceived temporal demand, and also RQ3 by showing that participants frequently attributed these judgments to uncertainty and interpretive ambiguity about system progress. They suggest that duration judgments appear sensitive to uncertainty and the availability of temporal cues, whereas perceived time pressure is shaped by how much usable recovery time a pause affords within the task context. These distinctions reinforce that temporal feedback affects the felt experience of waiting through several channels (uncertainty, pacing, and opportunities for recovery), none of which translated into measurable differences in performance in our study. Understanding these processes can help clarify when and why temporal feedback improves the experiential quality of waiting even when objective outcomes remain unchanged.

\subsection{\textbf{Insight 3:} \emph{Remaining-Time} feedback heightens negative affect more than \emph{Elapsed-Time}}
    
    Across all three Wait Durations, \emph{Remaining-Time} feedback elicited significantly higher frustration and significantly lower pleasantness than \emph{Elapsed-Time}, a key affective pattern for RQ2 that is consistent with participants’ accounts in RQ3. This aligns with prior work showing that time-remaining cues heighten temporal vigilance, impatience, and physiological arousal \cite{shalev_does_2013, komatsu_waiting_2024}. By continually signaling how much time remains, countdown-style information increases sensitivity to each passing second and elevates perceived pressure \cite{lallemand_enhancing_2012, zakay_role_1996}. In contrast, \emph{Elapsed-Time} feedback provides temporal orientation without emphasizing an impending endpoint. It conveys less anticipatory information and reduces preoccupation with the upcoming interval.
    
    \emph{Remaining-Time} serves an anticipatory role by updating users on how much time is left, whereas \emph{Elapsed-Time} provides temporal orientation without specifying what remains. Participants’ reflections illustrate this contrast. One participant in the \emph{Elapsed-Time} condition reported, “I was unsure of how much longer I’d wait, but had no effect on my performance,” while another participant in the \emph{Remaining-Time} condition described heightened arousal: “It made me more anxious and excited to begin… I was sorta stressed waiting.”
    
    The qualitative data support this distinction. Within the \emph{No Influence} theme, a larger share of coded instances came from the \emph{Elapsed-Time} condition (38.5\% of all participants) than from the \emph{Remaining-Time} condition (27.9\%). This suggests that temporal orientation without an endpoint indication was more frequently experienced as neutral or non-obtrusive. In contrast, the \emph{Affective Dysregulation} theme was more often associated with \emph{Remaining-Time} feedback (16.4\% of all participants, compared to 8.5\% in \emph{Elapsed-Time}), indicating that information framed as \emph{Remaining-Time} more readily elicited tension or impatience.

    Viewed together, these accounts suggest that the emotional costs are plausibly tied to the anticipatory content conveyed by \emph{Remaining-Time} information. This interpretation provides a complementary perspective to prior directionality work: rather than attributing affective responses solely to the visual act of counting down, our framing highlights that information about an approaching endpoint may itself shape affective experience. While our data cannot isolate which component of the representation is responsible, the patterns indicate that anticipatory emphasis is a promising explanatory lens for understanding why \emph{Remaining-Time} elicited higher frustration and lower pleasantness.

\subsection{\textbf{Insight 4:} Affective differences do not translate into differences in post-wait task performance}

Despite clear experiential differences across conditions, we did not observe corresponding differences in post-wait task performance. This directly addresses RQ1 by showing that neither Temporal Feedback Mode nor Wait Duration produced measurable differences in post-wait task performance in our setting, despite their effects on subjective experience (RQ2) and participants’ accounts (RQ3). Temporal Feedback Mode and Wait Duration shaped several subjective outcomes (including perceived duration, frustration, pleasantness, and temporal demand), and the qualitative themes revealed experiences ranging from restorative pause to affective dysregulation and interpretive ambiguity. Yet these experiential shifts did not impact how well participants performed on the subsequent task segment.

This pattern suggests a useful theoretical constraint. In our setting, temporal feedback appears to influence the felt quality of waiting without altering immediate task accuracy. This helps explain why prior work often reports strong effects of progress indicators on perceived waiting and satisfaction, but more modest or inconsistent effects on objective task measures \cite{nah_study_2004, myers_importance_1985, harrison_rethinking_2007}. From a design perspective, this distinction matters: improving the emotional experience of waiting does not guarantee improvements in performance. The primary benefits of temporal feedback may lie in perceived usability, trust, and willingness to continue using a system rather than in short-term task outcomes.

Seen in this light, our findings refine how progress indicators and countdowns should be understood. Much prior HCI work has emphasized their role in reducing the disruptive impact of delays. Our results suggest a more bounded picture: in relatively short, repetitive, low-complexity workflows like ours, temporal feedback functions chiefly as an experiential regulator rather than a performance-enhancing mechanism. Waiting is therefore not only an efficiency problem but also an experiential one shaped by attention, affect, expectations, and interpretations of system state. These contextual factors should guide how the results are generalized to other tasks.

\subsection{Boundary Conditions: Scope of Generalization}

Our findings arise from controlled, determinate waits in which the delay length was known in advance and presented through simple visual implementations of three informational modes: \emph{No Time Display}, \emph{Elapsed-Time}, and \emph{Remaining-Time}. Many real-world systems involve indeterminate waits whose durations cannot be predicted due to network variability or resource contention. The present results should therefore not be generalized to arbitrary or unpredictable delays. The mechanisms we observe, such as the influence of temporal cues on perceived duration, frustration, and uncertainty, may still operate in indeterminate scenarios, but the trade-offs likely change once duration estimates can be wrong or violated. In such settings, \emph{Remaining-Time} feedback risks eroding trust if a countdown completes prematurely, making \emph{Elapsed-Time} or \emph{No Time Display} paired with qualitative status cues (e.g., “still processing”) more appropriate defaults. When latency cannot be predicted, designers should avoid precise countdowns entirely and instead rely on qualitative status cues, conservative upper-bound estimates when available, and interfaces that allow users to redirect attention until the operation completes.

Generalization is also limited by the graphical form of temporal feedback. The same informational mode can be implemented through different visual metaphors, such as progress bars, loader animations, or spinners. Prior work shows that visual dynamics and salience modulate perceived duration and affect \cite{gomez_filled-duration_1979, myers_importance_1985, fang_effects_2022}, suggesting that richer animations may amplify or attenuate the experiential patterns observed here. Our results are best interpreted as characterizing how these informational categories shape waiting experience under a simple visual instantiation. Future work should systematically vary graphical form, animation style, and latency predictability to assess the robustness of these distinctions.

Task characteristics impose further boundary conditions. The present task was repetitive and quiz-like, which may explain why short pauses sometimes served as calming or restorative microbreaks. Brief externally imposed rests can restore attention during monotonous or demanding work \cite{ariga_brief_2011}. In contrast, interruptions during complex or problem-solving tasks can disrupt goal maintenance and increase frustration \cite{altmann_memory_2002}. Wait Duration also interacts with engagement. Short waits can function as effective microbreaks, whereas longer or unexpected waits risk boredom or disengagement \cite{mark_cost_2008}. Expectations shape perceived waiting as well; unexpected or unexplained delays typically feel longer and more frustrating than anticipated ones \cite{maister_psychology_1985}. These considerations underscore that effects of temporal feedback will vary across task types, delay structures, and user expectations.

\section{Design Guidelines}
\label{Design Guidelines}

Synthesizing these findings, we offer design considerations that reflect the trade-offs observed across Temporal Feedback Modes. Our quantitative results show that the three numerical modes primarily shaped the \emph{subjective experience} of waiting, while qualitative accounts clarify \emph{how} these experiential differences manifested. Because none of the modes improved post-wait performance, and because each mode carries distinct experiential benefits and drawbacks, the recommendations below are framed as context-sensitive considerations rather than prescriptive rules.

\begin{enumerate}
    \item \textbf{Reduce uncertainty when predictability is the primary user concern.} 
    
    Consider \emph{Remaining-Time} displays when the goal is to minimize uncertainty or provide reassurance during longer or potentially disruptive pauses. This mode consistently reduced the qualitative theme of interpretive ambiguity, but it also elevated frustration and lowered pleasantness. It is therefore most appropriate when the cost of ambiguity outweighs the emotional costs associated with increased temporal vigilance, such as in workflows where users must remain oriented or where delays might otherwise be interpreted as system failure.

    \item \textbf{Support emotional comfort through gentler temporal cues.}
    
    Use \emph{Elapsed-Time} displays when designers aim to maintain a calmer waiting experience. Participants reported lower frustration and higher pleasantness under this mode. In the qualitative analysis, it was described as less intrusive, reflected in the lowest incidence of \emph{Affective Dysregulation} codes. It was also characterized as more neutral, as indicated by the prevalence of statements that the wait had no meaningful influence on their experience. This suggests that \emph{Elapsed-Time} is a balanced default for systems where affective comfort or sustained engagement is more important than precise temporal predictability.

   \item \textbf{Temporal Feedback Modes shape experience more than performance.}
   
   In this study's context, affective differences across \emph{Remaining-Time}, \emph{Elapsed-Time}, and \emph{No Time Display} did not translate into post-wait performance differences. Designers therefore have the flexibility to select Temporal Feedback Modes based on experiential priorities (such as comfort, clarity, trust, or perceived pace), while attending to each mode’s trade-offs. For example, \emph{Remaining-Time} can reduce ambiguity but increase tension, elapsed-time offers a calmer experience, and \emph{No Time Display} reduces temporal vigilance but increases perceived duration and ambiguity.

\end{enumerate}

These findings derive from short, determinate waits (10–60 seconds) presented with simple numerical timers in a repetitive visual–spatial task. As aforementioned, effects may differ in settings with indeterminate latency, richer graphical forms (e.g., progress bars, loaders), or tasks with different cognitive and attentional demands. Designers should therefore interpret these recommendations as specific to contexts similar to our study.

The next section discusses the limitations of our study and highlights opportunities for future work, including the exploration of dynamic or adaptive feedback strategies to further optimize system-imposed wait experience across variable delay lengths.
\section{Limitations and Future Work}
\label{Limitations}

    \textbf{Task specificity and generalizability.}  
    Our study used a repetitive visual reasoning quiz task, which may make breaks feel more like a "restorative pause" than they would in non-repetitive or creative digital tasks. Wait periods might in other cases disrupt flow or increase frustration rather than support recovery. Moreover, in tasks like web search or AI chat, interruptions might lead to disengagement rather than reengagement \cite{arapakis_impact_2014}. Future studies should examine how different task types shape the thematic experiences of waits, particularly themes such as "Task Re-Engagement," "Cognitive Disruption," and "Interpretive Ambiguity," to validate and refine our insights across diverse application domains.
    
    \textbf{Role of time-related expectations.}  
    In our study, participants did not anticipate the wait periods; delays likely appeared unexpectedly and without explicit forewarning. In real-world contexts (e.g., downloading large files, software updates, AI-generated content), users often expect delays and adapt their behavior accordingly, such as by planning to multitask or shifting focus temporarily. This expectation can fundamentally reshape wait experiences.
    
    Prior research has also demonstrated that violations of wait expectations (i.e., when actual waits exceed or differ from what was anticipated) can significantly influence satisfaction, perceived duration, and perceived fairness, often more so than the objective length of the delay itself \cite{caruelle_clock_2023, whiting_closing_2009, maister_psychology_1985}.
    
    Future work could explicitly manipulate wait-time expectations (e.g., via pre-wait notifications or approximate duration cues) to examine how anticipatory frames interact with different temporal feedback modes, advancing our understanding of how expectation management moderates task performance and affective responses.

    \textbf{Granularity of wait durations and pacing.}  
    Our work focused on 10, 30, and 60 second delays, which capture many common system latency ranges but exclude much shorter microbreaks (<5s) and extended waits (>120s). We also did not examine intermediate intervals (e.g., 15s, 45s), where threshold effects or nonlinear transitions in user experience might emerge. Additionally, we employed a fixed 1Hz update frequency for both \emph{Remaining-Time} and \emph{Elapsed-Time} feedback modes, matching second-by-second progression. While prior research has explored how the speed and update dynamics of temporal feedback (e.g., progress bars) affect perceived duration, arousal, and satisfaction \cite{song_duration_2025, hurter_active_2012, cui_feedback_2022}, our study did not systematically investigate its impact on downstream task performance or cognitive and emotional outcomes. Future work could extend these insights by manipulating both duration granularity and feedback pacing in more diverse contexts to better understand trade-offs in performance, emotional regulation, and reengagement.
    
    \textbf{Directionality and informational completeness.} Conceptually, the two formats stand in opposition: count-up displays reveal the time that has elapsed, whereas count-down displays reveal the time remaining. Because a count-down progresses toward zero, it makes the remaining duration explicit and also conveys elapsed time implicitly. In practice, interfaces vary in how they present temporal information. Some show only elapsed time, others show only remaining time, while some provide both (e.g., “03/10 seconds”). Our study selected an elapsed-only count-up, and a remaining-only count-down, because these two formats are widely recognizable and represent common patterns in many contemporary systems. While these elements often co-occur in deployed interfaces, they are not logically inseparable. Fully disentangling them would require additional hybrids, such as a count-up display with an explicit endpoint or a count-down without a clear total duration (e.g., a timer that stops short of zero, or extends before reaching zero). Such extensions are valuable directions for future work aimed at isolating the pure effects of directionality or informational completeness. In the present study, however, our goal was to examine the experiential consequences of the three prevalent and ecologically familiar numerical feedback modes (\emph{No Time Display}, \emph{Elapsed-Time}, and \emph{Remaining-Time}), recognizing that they combine both directionality and information amount in ways that reflect many real-world wait screens.

    \textbf{Measurement of self-reported experience.} Our study relied on self-reported measures (PAD, NASA-TLX) and post-task qualitative feedback to infer affective and cognitive responses to delays. While informative, retrospective self-reports can be subject to recall biases and may not capture real-time fluctuations in arousal. Our recall-based items similarly capture memory of the wait rather than moment-to-moment perceived duration and may be influenced by attention, individual sensitivity to time, and exposure to temporal feedback. We interpret this variability as expected in retrospective duration judgments and consider these items as indicators of the wait’s salience rather than precise estimates of duration. Although prior work has proposed that countdowns can elevate self-reported arousal and affect subsequent time perception and decision-making \cite{shalev_does_2013, ghafurian_countdown_2020}, these effects have not been directly linked to physiological arousal. Future research should incorporate psychophysiological measures (e.g., galvanic skin response, heart-rate variability) to map moment-to-moment arousal dynamics more precisely and to clarify whether countdown-induced changes are mediated by physiological arousal or attentional mechanisms. Additionally, exploring how individual differences (e.g., time urgency, trait anxiety) moderate these responses would further enrich our understanding of user experience during system-imposed delays.

    \textbf{Sample characteristics and cultural considerations.}
    Our sample included participants based in the U.S., and they were recruited via Prolific, which may imply a generally higher level of digital familiarity. Prior work suggests that cultural norms and digital literacy can shape perceptions of waiting, patience thresholds, and trust in system feedback (e.g., differences in service expectations across cultural contexts \cite{maister_psychology_1985}). Future studies should examine temporal feedback responses in more diverse cultural and geographic settings, including non-Western contexts and individuals with lower digital literacy, to explore potential variability in wait-time experiences and preferences.

\vspace{0.2cm}

Overall, while our experimental design supports interpretations of the observed effects, real-world environments often involve multitasking, external interruptions, and richer media contexts. Embedding similar experiments into more ecologically valid settings (such as in collaborative, AI-based tools or mobile multitasking scenarios) would help clarify the practical boundaries of these insights and strengthen their applicability in everyday digital interactions.

\section{Conclusion}

A central contribution of this work is its evaluation of post-wait task performance across three temporal feedback modes (\emph{No Time Display}, \emph{Elapsed-Time}, and \emph{Remaining-Time}) within a fixed, system-imposed wait. Prior research on interruptions has shown that structuring when interruptions occur can support smoother resumption, but it has not examined how temporal feedback presented during a wait influences downstream performance. Our results show that although temporal feedback shapes users' perceived duration, frustration, and pleasantness, it does not produce measurable differences in post-wait task performance in this context. This divergence highlights an important condition for understanding temporal feedback in low-complexity, short, determinate waits: changes in the experience of waiting do not necessarily lead to changes in performance. Recognizing this distinction refines theoretical accounts of waiting and supports choosing feedback modes based on experiential considerations rather than presumed performance benefits.

More broadly, our results and design guidelines underscore that no single temporal-feedback mode is universally optimal. This is increasingly important as modern interactive systems rely on asynchronous computation and AI-driven processes, making waits a persistent part of the user experience. Our findings suggest that designing waits is less about influencing performance outcomes and more about shaping how users experience and interpret the system. This insight that invites deeper examination of the cognitive and affective processes that unfold during system-imposed waits.

\begin{acks}
 This work was supported by the National Science Foundation (NSF) under grants 1928614 and 2129076. We are also grateful to our colleagues for their constructive feedback.
\end{acks}

\bibliographystyle{ACM-Reference-Format}
\bibliography{references}

@article{ratwani_spatial_2008,
	title = {Spatial memory guides task resumption},
	volume = {16},
	issn = {1464-0716},
	doi = {10.1080/13506280802025791},
	 number = {8},
	journal = {Visual Cognition},
	author = {Ratwani, Raj M. and Trafton, J. Gregory},
	year = {2008},
	 
	 pages = {1001--1010},
	 }

@article{ariga_brief_2011,
	title = {Brief and rare mental "breaks" keep you focused: deactivation and reactivation of task goals preempt vigilance decrements},
	volume = {118},
	issn = {1873-7838},
	shorttitle = {Brief and rare mental "breaks" keep you focused},
	doi = {10.1016/j.cognition.2010.12.007},
	 language = {eng},
	number = {3},
	journal = {Cognition},
	author = {Ariga, Atsunori and Lleras, Alejandro},
	month = mar,
	year = {2011},
	pmid = {21211793},
	 pages = {439--443},
}

@article{bailey_need_2006,
	series = {Attention aware systems},
	title = {On the need for attention-aware systems: {Measuring} effects of interruption on task performance, error rate, and affective state},
	volume = {22},
	issn = {0747-5632},
	shorttitle = {On the need for attention-aware systems},
	url = {https://www.sciencedirect.com/science/article/pii/S074756320500107X},
	doi = {10.1016/j.chb.2005.12.009},
	 number = {4},
	urldate = {2025-08-10},
	journal = {Computers in Human Behavior},
	author = {Bailey, Brian P. and Konstan, Joseph A.},
	month = jul,
	year = {2006},
	 pages = {685--708},
	 }

@misc{nielsen_powers_2009,
	title = {Powers of 10: {Time} {Scales} in {User} {Experience}},
	shorttitle = {Powers of 10},
	url = {https://www.nngroup.com/articles/powers-of-10-time-scales-in-ux/},
	 language = {en},
	urldate = {2025-11-26},
	journal = {Nielsen Norman Group},
	author = {Nielsen, Jakob},
	month = oct,
	year = {2009},
	 }

@article{ramsay_psychological_1998,
	series = {{HCI} and {Information} {Retrieval}},
	title = {A psychological investigation of long retrieval times on the {World} {Wide} {Web}},
	volume = {10},
	issn = {0953-5438},
	url = {https://www.sciencedirect.com/science/article/pii/S0953543897000192},
	doi = {10.1016/S0953-5438(97)00019-2},
	 number = {1},
	urldate = {2025-11-26},
	journal = {Interacting with Computers},
	author = {Ramsay, Judith and Barbesi, Alessandro and Preece, Jenny},
	month = mar,
	year = {1998},
	 pages = {77--86},
	 }

@article{ballard_memory_1995,
	title = {Memory representations in natural tasks},
	volume = {7},
	issn = {0898-929X},
	doi = {10.1162/jocn.1995.7.1.66},
	 language = {eng},
	number = {1},
	journal = {Journal of Cognitive Neuroscience},
	author = {Ballard, D. H. and Hayhoe, M. M. and Pelz, J. B.},
	year = {1995},
	pmid = {23961754},
	pages = {66--80},
}

@inproceedings{liu_supporting_2014,
	address = {New York, NY, USA},
	series = {{CSCW} '14},
	title = {Supporting task resumption using visual feedback},
	isbn = {978-1-4503-2540-0},
	url = {https://dl.acm.org/doi/10.1145/2531602.2531710},
	doi = {10.1145/2531602.2531710},
	 urldate = {2025-11-26},
	booktitle = {Proceedings of the 17th {ACM} conference on {Computer} supported cooperative work \& social computing},
	publisher = {Association for Computing Machinery},
	author = {Liu, Yikun and Jia, Yuan and Pan, Wei and Pfaff, Mark S.},
	month = feb,
	year = {2014},
	pages = {767--777},
	 }

@article{chen_effects_2025,
	title = {The effects of cues on task interruption recovery in a concurrent multitasking environment},
	volume = {15},
	copyright = {2025 The Author(s)},
	issn = {2045-2322},
	url = {https://www.nature.com/articles/s41598-025-09358-4},
	doi = {10.1038/s41598-025-09358-4},
	 language = {en},
	number = {1},
	urldate = {2025-11-26},
	journal = {Scientific Reports},
	author = {Chen, Yueyuan and Zhang, Chuanwang and Fang, Weining and Ma, Jing},
	month = jul,
	year = {2025},
	  pages = {25992},
	 }

@article{miller_misunderstanding_2001,
	title = {Misunderstanding analysis of covariance},
	volume = {110},
	issn = {0021-843X},
	doi = {10.1037//0021-843x.110.1.40},
	 language = {eng},
	number = {1},
	journal = {Journal of Abnormal Psychology},
	author = {Miller, G. A. and Chapman, J. P.},
	month = feb,
	year = {2001},
	pmid = {11261398},
	 pages = {40--48},
}

@misc{lenth_emmeans_2025,
	title = {emmeans: {Estimated} {Marginal} {Means}, aka {Least}-{Squares} {Means}},
	copyright = {GPL-2 {\textbar} GPL-3},
	shorttitle = {emmeans},
	url = {https://cran.r-project.org/web/packages/emmeans/index.html},
	urldate = {2025-08-12},
	author = {Lenth, Russell V. and Banfai, Balazs and Bolker, Ben and Buerkner, Paul and Giné-Vázquez, Iago and Herve, Maxime and Jung, Maarten and Love, Jonathon and Miguez, Fernando and Piaskowski, Julia and Riebl, Hannes and Singmann, Henrik},
	month = jul,
	year = {2025},
}

@article{liu_effects_2014,
	title = {The {Effects} of {Interactive} {Latency} on {Exploratory} {Visual} {Analysis}},
	volume = {20},
	issn = {1941-0506},
	url = {https://ieeexplore.ieee.org/document/6876022},
	doi = {10.1109/TVCG.2014.2346452},
	 number = {12},
	urldate = {2025-08-10},
	journal = {IEEE Transactions on Visualization and Computer Graphics},
	author = {Liu, Zhicheng and Heer, Jeffrey},
	month = dec,
	year = {2014},
	 pages = {2122--2131},
}

@inproceedings{adamczyk_if_2004,
	address = {New York, NY, USA},
	series = {{CHI} '04},
	title = {If not now, when? the effects of interruption at different moments within task execution},
	isbn = {978-1-58113-702-6},
	shorttitle = {If not now, when?},
	url = {https://dl.acm.org/doi/10.1145/985692.985727},
	doi = {10.1145/985692.985727},
	 urldate = {2025-08-10},
	booktitle = {Proceedings of the {SIGCHI} {Conference} on {Human} {Factors} in {Computing} {Systems}},
	publisher = {Association for Computing Machinery},
	author = {Adamczyk, Piotr D. and Bailey, Brian P.},
	month = apr,
	year = {2004},
	pages = {271--278},
}

@article{altmann_memory_2002,
	title = {Memory for goals: an activation-based model},
	volume = {26},
	issn = {0364-0213},
	shorttitle = {Memory for goals},
	url = {https://www.sciencedirect.com/science/article/pii/S0364021301000581},
	doi = {10.1016/S0364-0213(01)00058-1},
	 number = {1},
	urldate = {2025-08-10},
	journal = {Cognitive Science},
	author = {Altmann, Erik M and Trafton, J. Gregory},
	month = jan,
	year = {2002},
	 pages = {39--83},
}

@inproceedings{harrison_faster_2010,
	address = {New York, NY, USA},
	series = {{CHI} '10},
	title = {Faster progress bars: manipulating perceived duration with visual augmentations},
	isbn = {978-1-60558-929-9},
	shorttitle = {Faster progress bars},
	url = {https://dl.acm.org/doi/10.1145/1753326.1753556},
	doi = {10.1145/1753326.1753556},
	 urldate = {2025-08-10},
	booktitle = {Proceedings of the {SIGCHI} {Conference} on {Human} {Factors} in {Computing} {Systems}},
	publisher = {Association for Computing Machinery},
	author = {Harrison, Chris and Yeo, Zhiquan and Hudson, Scott E.},
	month = apr,
	year = {2010},
	pages = {1545--1548},
}

@article{dellaert_how_1999,
	title = {How tolerable is delay?: {Consumers}’ evaluations of internet web sites after waiting},
	volume = {13},
	issn = {1094-9968},
	shorttitle = {How tolerable is delay?},
	url = {https://www.sciencedirect.com/science/article/pii/S1094996899702252},
	doi = {10.1002/(SICI)1520-6653(199924)13:1<41::AID-DIR4>3.0.CO;2-S},
	 number = {1},
	urldate = {2025-08-10},
	journal = {Journal of Interactive Marketing},
	author = {Dellaert, Benedict G. C. and Kahn, Barbara E.},
	month = jan,
	year = {1999},
	pages = {41--54},
}

@inproceedings{lee_while_2025,
	address = {New York, NY, USA},
	series = {{CHI} {EA} '25},
	title = {While {We} {Wait}... {How} {Users} {Perceive} {Waiting} {Times} and {Generation} {Cues} during {AI} {Image} {Generation}},
	isbn = {9798400713958},
	url = {https://doi.org/10.1145/3706599.3719725},
	doi = {10.1145/3706599.3719725},
	 urldate = {2025-08-10},
	booktitle = {Proceedings of the {Extended} {Abstracts} of the {CHI} {Conference} on {Human} {Factors} in {Computing} {Systems}},
	publisher = {Association for Computing Machinery},
	author = {Lee, Hui Min and Yadav, Davis and Lee, Sangwook and Govindarazan, Keerthana and Chen, Cheng and Sundar, S. Shyam},
	month = apr,
	year = {2025},
	pages = {1--8},
}

@article{hong_when_2013,
	title = {When {Filling} the {Wait} {Makes} it {Feel} {Longer}: {A} {Paradigm} {Shift} {Perspective} for {Managing} {Online} {Delay}},
	volume = {37},
	issn = {0276-7783},
	shorttitle = {When {Filling} the {Wait} {Makes} it {Feel} {Longer}},
	url = {https://www.jstor.org/stable/43825915},
	 number = {2},
	urldate = {2025-08-04},
	journal = {MIS Quarterly},
	author = {Hong, Weiyin and Hess, Traci J. and Hardin, Andrew},
	year = {2013},
	 pages = {383--406},
}

@article{galletta_when_2006,
	title = {When the {Wait} {Isnt} {So} {Bad}: {The} {Interacting} {Effects} of {Website} {Delay}, {Familiarity}, and {Breadth}},
	volume = {17},
	issn = {1526-5536},
	shorttitle = {When the {Wait} {Isnt} {So} {Bad}},
	url = {https://doi.org/10.1287/isre.1050.0073},
	doi = {10.1287/isre.1050.0073},
	 number = {1},
	urldate = {2025-08-04},
	journal = {Info. Sys. Research},
	author = {Galletta, Dennis F. and Henry, Raymond M. and McCoy, Scott and Polak, Peter},
	month = mar,
	year = {2006},
	pages = {20--37},
}

@article{hoxmeier_system_2000,
	title = {System {Response} {Time} and {User} {Satisfaction}: {An} {Experimental} {Study} of {Browser}-based {Applications}},
	shorttitle = {System {Response} {Time} and {User} {Satisfaction}},
	url = {https://aisel.aisnet.org/amcis2000/347},
	journal = {AMCIS 2000 Proceedings},
	author = {Hoxmeier, John and DiCesare, Chris},
	month = jan,
	year = {2000},
}

@article{whiting_closing_2009,
	title = {Closing the gap between perceived and actual waiting times in a call center: results from a field study},
	volume = {23},
	issn = {0887-6045},
	shorttitle = {Closing the gap between perceived and actual waiting times in a call center},
	url = {https://doi.org/10.1108/08876040910973396},
	doi = {10.1108/08876040910973396},
	 number = {5},
	urldate = {2025-07-09},
	journal = {Journal of Services Marketing},
	author = {Whiting, Anita and Donthu, Naveen},
	editor = {Mukherjee, Avinandan and Malhotra, Neeru},
	month = jan,
	year = {2009},
	  pages = {279--288},
}

@article{caruelle_clock_2023,
	title = {The clock is ticking—{Or} is it? {Customer} satisfaction response to waiting shorter vs. longer than expected during a service encounter},
	volume = {99},
	issn = {0022-4359},
	shorttitle = {The clock is ticking—{Or} is it?},
	url = {https://www.sciencedirect.com/science/article/pii/S0022435923000143},
	doi = {10.1016/j.jretai.2023.03.003},
	 number = {2},
	urldate = {2025-07-09},
	journal = {Journal of Retailing},
	author = {Caruelle, Delphine and Lervik-Olsen, Line and Gustafsson, Anders},
	month = jun,
	year = {2023},
	 pages = {247--264},
}

@article{maister_psychology_1985,
	title = {The {Psychology} of {Waiting} {Lines}},
	volume = {In J.A. Czepiel, M.R. Solomon, \& C. Surprenant (Eds.), The Service Encounter},
	journal = {Lexington: Lexington Books},
	author = {Maister, D H},
	year = {1985},
	pages = {113--123},
}

@article{dabrowski_40years_2011,
	title = {40years of searching for the best computer system response time},
	volume = {23},
	issn = {0953-5438},
	url = {https://doi.org/10.1016/j.intcom.2011.05.008},
	doi = {10.1016/j.intcom.2011.05.008},
	 number = {5},
	urldate = {2025-06-03},
	journal = {Interact. Comput.},
	author = {Dabrowski, Jim and Munson, Ethan V.},
	month = sep,
	year = {2011},
	pages = {555--564},
}

@inproceedings{boucsein_forty_2009,
	address = {Berlin, Heidelberg},
	title = {Forty {Years} of {Research} on {System} {Response} {Times} – {What} {Did} {We} {Learn} from {It}?},
	isbn = {978-3-642-01293-8},
	doi = {10.1007/978-3-642-01293-8_42},
	 language = {en},
	booktitle = {Industrial {Engineering} and {Ergonomics}},
	publisher = {Springer},
	author = {Boucsein, Wolfram},
	editor = {Schlick, Christopher M.},
	year = {2009},
	pages = {575--593},
}

@inproceedings{attig_system_2017,
	address = {Cham},
	title = {System {Latency} {Guidelines} {Then} and {Now} – {Is} {Zero} {Latency} {Really} {Considered} {Necessary}?},
	isbn = {978-3-319-58475-1},
	doi = {10.1007/978-3-319-58475-1_1},
	 language = {en},
	booktitle = {Engineering {Psychology} and {Cognitive} {Ergonomics}: {Cognition} and {Design}},
	publisher = {Springer International Publishing},
	author = {Attig, Christiane and Rauh, Nadine and Franke, Thomas and Krems, Josef F.},
	editor = {Harris, Don},
	year = {2017},
	 pages = {3--14},
}

@article{song_duration_2025,
	title = {Duration and {Speed} {Perception} of {Loading} {Processes}: {Influence} of {Progress} {Bar}’s {Physical} and {Symbolic} {Speed}},
	volume = {0},
	issn = {1044-7318},
	shorttitle = {Duration and {Speed} {Perception} of {Loading} {Processes}},
	url = {https://doi.org/10.1080/10447318.2024.2446501},
	doi = {10.1080/10447318.2024.2446501},
	 number = {0},
	urldate = {2025-05-31},
	journal = {International Journal of Human–Computer Interaction},
	author = {Song, Yuanming and , Shuang, Liang and , Baolin, Li and , Guojie, Ma and and Zhuang, Xiangling},
	year = {2025},
	 
	 pages = {1--11},
}

@inproceedings{harrison_rethinking_2007,
	address = {New York, NY, USA},
	series = {{UIST} '07},
	title = {Rethinking the progress bar},
	isbn = {978-1-59593-679-0},
	url = {https://dl.acm.org/doi/10.1145/1294211.1294231},
	doi = {10.1145/1294211.1294231},
	 urldate = {2025-05-31},
	booktitle = {Proceedings of the 20th annual {ACM} symposium on {User} interface software and technology},
	publisher = {Association for Computing Machinery},
	author = {Harrison, Chris and Amento, Brian and Kuznetsov, Stacey and Bell, Robert},
	month = oct,
	year = {2007},
	pages = {115--118},
}

@inproceedings{hurter_active_2012,
	address = {Swindon, GBR},
	series = {{BCS}-{HCI} '12},
	title = {Active progress bar: aiding the switch to temporary activities},
	shorttitle = {Active progress bar},
	 urldate = {2025-05-31},
	booktitle = {Proceedings of the 26th {Annual} {BCS} {Interaction} {Specialist} {Group} {Conference} on {People} and {Computers}},
	publisher = {BCS Learning \& Development Ltd.},
	author = {Hurter, Christophe and Cowan, Benjamin R. and Girouard, Audrey and Riche, Nathalie Henry},
	month = sep,
	year = {2012},
	pages = {99--108},
}

@article{ozel_effect_2004,
	title = {Effect of arousal on internal clock speed in real action and mental imagery},
	volume = {58},
	issn = {1196-1961},
	doi = {10.1037/h0087444},
	 language = {eng},
	number = {3},
	journal = {Canadian Journal of Experimental Psychology = Revue Canadienne De Psychologie Experimentale},
	author = {Ozel, Sylvie and Larue, Jacques and Dosseville, Fabrice},
	month = sep,
	year = {2004},
	pmid = {15487439},
	 pages = {196--205},
}

@article{droit-volet_emotional_2016,
	title = {The emotional body and time perception},
	volume = {30},
	issn = {0269-9931},
	url = {https://doi.org/10.1080/02699931.2015.1023180},
	doi = {10.1080/02699931.2015.1023180},
	 number = {4},
	urldate = {2025-05-31},
	journal = {Cognition and Emotion},
	author = {Droit-Volet, Sylvie and and Gil, Sandrine},
	month = may,
	year = {2016},
	pmid = {25817441},
	 
	 pages = {687--699},
}

@article{boltz_changes_1994,
	title = {Changes in internal tempo and effects on the learning and remembering of event durations},
	volume = {20},
	issn = {1939-1285},
	doi = {10.1037/0278-7393.20.5.1154},
	 number = {5},
	journal = {Journal of Experimental Psychology: Learning, Memory, and Cognition},
	author = {Boltz, Marilyn G.},
	year = {1994},
	 
	 pages = {1154--1171},
}

@article{reed_use_2004,
	title = {Use of temporal and spatial information in estimating event completion time},
	volume = {32},
	issn = {1532-5946},
	url = {https://doi.org/10.3758/BF03196858},
	doi = {10.3758/BF03196858},
	 language = {en},
	number = {2},
	urldate = {2025-05-31},
	journal = {Memory \& Cognition},
	author = {Reed, Stephen K. and Hoffman, Bob},
	month = mar,
	year = {2004},
	 pages = {271--282},
}

@article{odonnell_how_1996,
	title = {How machine delays change user strategies},
	volume = {28},
	issn = {0736-6906},
	url = {https://dl.acm.org/doi/10.1145/226650.226665},
	doi = {10.1145/226650.226665},
	 number = {2},
	urldate = {2025-05-31},
	journal = {SIGCHI Bull.},
	author = {O'Donnell, Paddy and Draper, Stephen W.},
	month = apr,
	year = {1996},
	pages = {39--42},
}

@article{shalev_does_2013,
	title = {Does time fly when you're counting down? {The} effect of counting direction on subjective time judgment},
	volume = {23},
	issn = {1532-7663},
	shorttitle = {Does time fly when you're counting down?},
	doi = {10.1016/j.jcps.2012.08.002},
	 number = {2},
	journal = {Journal of Consumer Psychology},
	author = {Shalev, Edith and Morwitz, Vicki G.},
	year = {2013},
	 
	 pages = {220--227},
}

@article{gibbon_scalar_1984,
	title = {Scalar {Timing} in {Memory}},
	volume = {423},
	issn = {1749-6632},
	url = {https://onlinelibrary.wiley.com/doi/abs/10.1111/j.1749-6632.1984.tb23417.x},
	doi = {10.1111/j.1749-6632.1984.tb23417.x},
	language = {en},
	number = {1},
	urldate = {2025-05-21},
	journal = {Annals of the New York Academy of Sciences},
	author = {Gibbon, John and Church, Russell M. and Meck, Warren H.},
	year = {1984},
	 pages = {52--77},
}

@article{treisman_temporal_1963,
	title = {Temporal discrimination and the indifference interval: {Implications} for a model of the "internal clock"},
	volume = {77},
	issn = {0096-9753},
	shorttitle = {Temporal discrimination and the indifference interval},
	doi = {10.1037/h0093864},
	 number = {13},
	journal = {Psychological Monographs: General and Applied},
	author = {Treisman, Michel},
	year = {1963},
	 
	 pages = {1--31},
}

@incollection{zakay_role_1996,
	address = {Amsterdam, Netherlands},
	series = {Advances in psychology, {Vol}. 115},
	title = {The role of attention in time estimation processes},
	isbn = {978-0-444-82114-0},
	 booktitle = {Time, internal clocks and movement},
	publisher = {North-Holland/Elsevier Science Publishers},
	author = {Zakay, Dan and Block, Richard A.},
	year = {1996},
	doi = {10.1016/S0166-4115(96)80057-4},
	 pages = {143--164},
}

@inproceedings{wobbrock_aligned_2011,
	address = {Vancouver BC Canada},
	title = {The aligned rank transform for nonparametric factorial analyses using only anova procedures},
	isbn = {978-1-4503-0228-9},
	url = {https://dl.acm.org/doi/10.1145/1978942.1978963},
	doi = {10.1145/1978942.1978963},
	 language = {en},
	urldate = {2025-04-20},
	booktitle = {Proceedings of the {SIGCHI} {Conference} on {Human} {Factors} in {Computing} {Systems}},
	publisher = {ACM},
	author = {Wobbrock, Jacob O. and Findlater, Leah and Gergle, Darren and Higgins, James J.},
	month = may,
	year = {2011},
	pages = {143--146},
}

@article{braun_using_2006,
	title = {Using thematic analysis in psychology},
	volume = {3},
	issn = {1478-0895},
	doi = {10.1191/1478088706qp063oa},
	 number = {2},
	journal = {Qualitative Research in Psychology},
	author = {Braun, Virginia and Clarke, Victoria},
	year = {2006},
	 
	 pages = {77--101},
}

@incollection{hart_development_1988,
	series = {Human {Mental} {Workload}},
	title = {Development of {NASA}-{TLX} ({Task} {Load} {Index}): {Results} of {Empirical} and {Theoretical} {Research}},
	volume = {52},
	shorttitle = {Development of {NASA}-{TLX} ({Task} {Load} {Index})},
	url = {https://www.sciencedirect.com/science/article/pii/S0166411508623869},
	 urldate = {2025-04-20},
	booktitle = {Advances in {Psychology}},
	publisher = {North-Holland},
	author = {Hart, Sandra G. and Staveland, Lowell E.},
	editor = {Hancock, Peter A. and Meshkati, Najmedin},
	month = jan,
	year = {1988},
	doi = {10.1016/S0166-4115(08)62386-9},
	pages = {139--183},
}

@article{selvidge_how_1999,
	title = {How {Long} is {Too} {Long} to {Wait} for a {Website} to {Load}?},
	number = {1.2},
	journal = {Usability News},
	author = {Selvidge, Paula},
	year = {1999},
	pages = {1999--2001},
}

@article{fang_effects_2022,
	title = {Effects of {Different} {Visual} {Feedback} {Types} on {Perception} of {Online} {Wait}},
	volume = {39},
	url = {https://www.iieta.org/journals/ts/paper/10.18280/ts.390423},
	doi = {https://doi.org/10.18280/ts.390423},
	language = {en},
	number = {4},
	urldate = {2025-04-19},
	journal = {Traitement du Signal},
	author = {Fang, N.L. and Hu, T. and Shi, M.D. and Liu, Z.H.},
	year = {2022},
	pages = {1303--1312},
}

@article{ghafurian_countdown_2020,
	title = {Countdown timer speed: {A} trade-off between delay duration perception and recall},
	volume = {27},
	issn = {1557-7325},
	shorttitle = {Countdown timer speed},
	doi = {10.1145/3380961},
	 number = {2},
	journal = {ACM Transactions on Computer-Human Interaction},
	author = {Ghafurian, Moojan and Reitter, David and Ritter, Frank E.},
	year = {2020},
	 
	 pages = {1--25},
}

@article{wickens_multiple_2008,
	title = {Multiple resources and mental workload},
	volume = {50},
	issn = {0018-7208},
	doi = {10.1518/001872008X288394},
	 
	language = {eng},
	number = {3},
	journal = {Human Factors},
	author = {Wickens, Christopher D.},
	month = jun,
	year = {2008},
	pmid = {18689052},
	 pages = {449--455},
}

@article{wang_effect_2021,
	title = {The effect of mobile applications’ initial loading pages on users’ mental state and behavior},
	volume = {68},
	issn = {0141-9382},
	url = {https://www.sciencedirect.com/science/article/pii/S0141938221000214},
	doi = {10.1016/j.displa.2021.102007},
	 urldate = {2024-10-07},
	journal = {Displays},
	author = {Wang, Yuzhen and Huang, Yanqun and Li, Jutao and Zhang, Jie},
	month = jul,
	year = {2021},
	 pages = {102007},
}

@article{trafton_task_2007,
	title = {Task {Interruptions}},
	volume = {3},
	issn = {1557-234X},
	url = {https://doi.org/10.1518/155723408X299852},
	doi = {10.1518/155723408X299852},
	 language = {en},
	number = {1},
	urldate = {2024-11-17},
	journal = {Reviews of Human Factors and Ergonomics},
	author = {Trafton, J. Gregory and Monk, Christopher A.},
	month = nov,
	year = {2007},
	 pages = {111--126},
}

@article{nah_study_2004,
	title = {A study on tolerable waiting time: how long are {Web} users willing to wait?},
	volume = {23},
	issn = {0144-929X},
	shorttitle = {A study on tolerable waiting time},
	url = {https://doi.org/10.1080/01449290410001669914},
	doi = {10.1080/01449290410001669914},
	 number = {3},
	urldate = {2024-11-06},
	journal = {Behaviour \& Information Technology},
	author = {Nah, Fiona Fui-Hoon},
	month = may,
	year = {2004},
	 
	pages = {153--163},
}

@article{myers_importance_1985,
	title = {The importance of percent-done progress indicators for computer-human interfaces},
	volume = {16},
	issn = {0736-6906},
	url = {https://dl.acm.org/doi/10.1145/1165385.317459},
	doi = {10.1145/1165385.317459},
	 number = {4},
	urldate = {2024-10-17},
	journal = {SIGCHI Bull.},
	author = {Myers, Brad A.},
	month = apr,
	year = {1985},
	pages = {11--17},
}

@article{monk_effect_2008,
	title = {The effect of interruption duration and demand on resuming suspended goals},
	volume = {14},
	issn = {1939-2192},
	doi = {10.1037/a0014402},
	 number = {4},
	journal = {Journal of Experimental Psychology: Applied},
	author = {Monk, Christopher A. and Trafton, J. Gregory and Boehm-Davis, Deborah A.},
	year = {2008},
	 
	 pages = {299--313},
}

@inproceedings{mark_cost_2008,
	address = {Florence Italy},
	title = {The cost of interrupted work: more speed and stress},
	isbn = {978-1-60558-011-1},
	shorttitle = {The cost of interrupted work},
	url = {https://dl.acm.org/doi/10.1145/1357054.1357072},
	doi = {10.1145/1357054.1357072},
	 language = {en},
	urldate = {2024-11-24},
	booktitle = {Proceedings of the {SIGCHI} {Conference} on {Human} {Factors} in {Computing} {Systems}},
	publisher = {ACM},
	author = {Mark, Gloria and Gudith, Daniela and Klocke, Ulrich},
	month = apr,
	year = {2008},
	pages = {107--110},
}

@inproceedings{lallemand_enhancing_2012,
	address = {New York, NY, USA},
	series = {{DIS} '12},
	title = {Enhancing {User} {eXperience} during waiting time in {HCI}: contributions of cognitive psychology},
	isbn = {978-1-4503-1210-3},
	shorttitle = {Enhancing {User} {eXperience} during waiting time in {HCI}},
	url = {https://dl.acm.org/doi/10.1145/2317956.2318069},
	doi = {10.1145/2317956.2318069},
	 urldate = {2024-10-17},
	booktitle = {Proceedings of the {Designing} {Interactive} {Systems} {Conference}},
	publisher = {Association for Computing Machinery},
	author = {Lallemand, Carine and Gronier, Guillaume},
	month = jun,
	year = {2012},
	pages = {751--760},
}

@article{mehrabian_pleasure-arousal-dominance_1996,
	title = {Pleasure-arousal-dominance: {A} general framework for describing and measuring individual differences in temperament},
	volume = {14},
	issn = {1936-4733},
	shorttitle = {Pleasure-arousal-dominance},
	doi = {10.1007/BF02686918},
	 number = {4},
	journal = {Current Psychology: A Journal for Diverse Perspectives on Diverse Psychological Issues},
	author = {Mehrabian, Albert},
	year = {1996},
	 
	 pages = {261--292},
}

@inproceedings{komatsu_waiting_2024,
	address = {New York, NY, USA},
	series = {{CHI} '24},
	title = {Waiting {Time} {Perceptions} for {Faster} {Count}-downs/ups {Are} {More} {Sensitive} {Than} {Slower} {Ones}: {Experimental} {Investigation} and {Its} {Application}},
	isbn = {9798400703300},
	shorttitle = {Waiting {Time} {Perceptions} for {Faster} {Count}-downs/ups {Are} {More} {Sensitive} {Than} {Slower} {Ones}},
	url = {https://doi.org/10.1145/3613904.3641942},
	doi = {10.1145/3613904.3641942},
	 urldate = {2024-10-02},
	booktitle = {Proceedings of the 2024 {CHI} {Conference} on {Human} {Factors} in {Computing} {Systems}},
	publisher = {Association for Computing Machinery},
	author = {Komatsu, Takanori and Xie, Chenxi and Yamada, Seiji},
	month = may,
	year = {2024},
	pages = {1--13},
}

@inproceedings{johnson_clevr_2017,
	title = {{CLEVR}: {A} {Diagnostic} {Dataset} for {Compositional} {Language} and {Elementary} {Visual} {Reasoning}},
	shorttitle = {{CLEVR}},
	url = {https://ieeexplore.ieee.org/document/8099698},
	doi = {10.1109/CVPR.2017.215},
	 urldate = {2025-01-13},
	booktitle = {2017 {IEEE} {Conference} on {Computer} {Vision} and {Pattern} {Recognition} ({CVPR})},
	author = {Johnson, Justin and Hariharan, Bharath and van der Maaten, Laurens and Fei-Fei, Li and Zitnick, C. Lawrence and Girshick, Ross},
	month = jul,
	year = {2017},
	  pages = {1988--1997},
}

@article{haigh_role_2021,
	title = {The role of {Weber}’s law in human time perception},
	volume = {83},
	issn = {1943-393X},
	url = {https://doi.org/10.3758/s13414-020-02128-6},
	doi = {10.3758/s13414-020-02128-6},
	 language = {en},
	number = {1},
	urldate = {2024-10-07},
	journal = {Attention, Perception, \& Psychophysics},
	author = {Haigh, Andrew and Apthorp, Deborah and Bizo, Lewis A.},
	month = jan,
	year = {2021},
	 pages = {435--447},
}

@article{gomez_filled-duration_1979,
	title = {The filled-duration illusion: {The} function of temporal and nontemporal set},
	volume = {25},
	issn = {1532-5962},
	shorttitle = {The filled-duration illusion},
	url = {https://doi.org/10.3758/BF03199853},
	doi = {10.3758/BF03199853},
	 language = {en},
	number = {5},
	urldate = {2024-10-07},
	journal = {Perception \& Psychophysics},
	author = {Gomez, Louis M. and Robertson, Lynn C.},
	month = sep,
	year = {1979},
	 pages = {432--438},
}

@inproceedings{cutrell_effects_2000,
	address = {New York, NY, USA},
	series = {{CHI} {EA} '00},
	title = {Effects of instant messaging interruptions on computing tasks},
	isbn = {978-1-58113-248-9},
	url = {https://dl.acm.org/doi/10.1145/633292.633351},
	doi = {10.1145/633292.633351},
	 urldate = {2024-11-24},
	booktitle = {{CHI} '00 {Extended} {Abstracts} on {Human} {Factors} in {Computing} {Systems}},
	publisher = {Association for Computing Machinery},
	author = {Cutrell, Edward B. and Czerwinski, Mary and Horvitz, Eric},
	month = apr,
	year = {2000},
	pages = {99--100},
}

@article{cui_feedback_2022,
	title = {Feedback frequency of graphical countdown timers affects pedestrian estimated waiting time at red lights},
	volume = {88},
	issn = {1369-8478},
	url = {https://www.sciencedirect.com/science/article/pii/S1369847822001012},
	doi = {10.1016/j.trf.2022.05.015},
	 urldate = {2024-11-17},
	journal = {Transportation Research Part F: Traffic Psychology and Behaviour},
	author = {Cui, Zixin and Zhuang, Xiangling and Chen, Wenxiang and Ma, Guojie},
	month = jul,
	year = {2022},
	 pages = {184--196},
}

@article{block_prospective_1997,
	title = {Prospective and retrospective duration judgments: {A} meta-analytic review},
	volume = {4},
	issn = {1531-5320},
	shorttitle = {Prospective and retrospective duration judgments},
	url = {https://doi.org/10.3758/BF03209393},
	doi = {10.3758/BF03209393},
	 language = {en},
	number = {2},
	urldate = {2024-11-17},
	journal = {Psychonomic Bulletin \& Review},
	author = {Block, Richard A. and Zakay, Dan},
	month = jun,
	year = {1997},
	 pages = {184--197},
}

@inproceedings{arapakis_impact_2014,
	address = {New York, NY, USA},
	series = {{SIGIR} '14},
	title = {Impact of response latency on user behavior in web search},
	isbn = {978-1-4503-2257-7},
	url = {https://doi.org/10.1145/2600428.2609627},
	doi = {10.1145/2600428.2609627},
	 urldate = {2025-01-24},
	booktitle = {Proceedings of the 37th international {ACM} {SIGIR} conference on {Research} \& development in information retrieval},
	publisher = {Association for Computing Machinery},
	author = {Arapakis, Ioannis and Bai, Xiao and Cambazoglu, B. Barla},
	month = jul,
	year = {2014},
	pages = {103--112},
}

\appendix
\section{Examples of Task Questions}
\label{appendix-examples}

\begin{figure}[H]

\textbf{Example 1.} There is a purple cube; what number of large spheres are
    to the left of it?\\[4pt]
    \includegraphics[width=1\linewidth]{Figures/CLEVR_val_001167.png}\\[10pt]
    
    \textbf{Example 2.} How many other spheres are made of the same material
    as the small yellow sphere?\\[4pt]
    \includegraphics[width=1\linewidth]{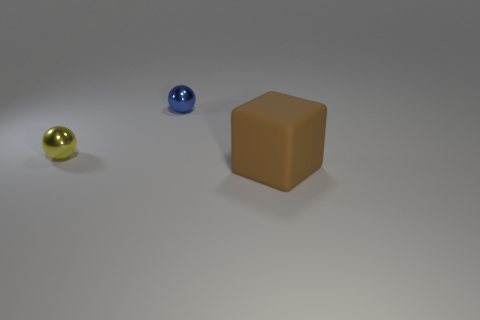}\\[6pt]
    
    \textbf{Answer options (choose one):}\\
    0 \quad 1 \quad 2 \quad 3 \quad 4

    \Description{Two example CLEVR-style visual reasoning stimuli. Example 1 shows multiple colored 3D objects, including cubes and spheres of different sizes and materials, arranged on a flat surface for a left–right spatial counting task. Example 2 shows a small yellow sphere and other objects of varying materials for a material-matching counting task. Answer choices are integers from 0 to 4.}

\end{figure}

\section{Qualitative Codebook for Free‐Response Items}
\label{appendix-qualitative codebook}

\vspace{4mm}
Below are the operational definitions used to code responses for each of the seven themes discussed in Section~\ref{Results - Qual}. 

\begin{table}[H]
    \centering
    \Description{Qualitative codebook listing seven codes used to analyze free-response items, with definitions and inclusion criteria. Codes: Perceived Irrelevance — wait had no measurable effect on emotion, attention, motivation, or performance; Restorative Pause — mental break or recovery; Task Re-Engagement — planning, strategy, or motivation for the next segment; Affective Regulation — mood calmed or anxiety eased; Affective Dysregulation — stress, frustration, anxiety, or impatience caused or intensified by the wait; Cognitive Disruption — interrupted attention or flow, harder to resume; Interpretive Ambiguity — uncertainty about system status or why the wait occurred.}
    \label{tab:appendix-free-response-codebook}
    \begin{tabular}{p{3cm} p{5.8cm}}
    \toprule
    \textbf{Code} & \textbf{Definition / Inclusion Criteria} \\
    \midrule
    Perceived Irrelevance &
    Any comment stating or implying that the wait had no measurable influence on emotion, attention, motivation, or performance. \\
    \addlinespace[4pt]

    Restorative Pause &
    Any description of the wait serving as a mental break, reducing fatigue or tension, or providing recovery. \\
    \addlinespace[4pt]

    Task Re‐Engagement &
    Any comment indicating the wait was used to plan, strategize, or boost motivation for the upcoming segment. \\
    \addlinespace[4pt]

    Affective Regulation &
    Any mention that the wait helped regulate mood or emotions positively, such as easing anxiety or feeling soothed. \\
    \addlinespace[4pt]

    Affective Dysregulation &
    Any expression of negative arousal, such as stress, frustration, anxiety, or impatience, directly caused or intensified by the wait. \\
    \addlinespace[4pt]

    Cognitive Disruption &
    Any comment indicating that the wait interrupted attention, flow, or focus, making it harder to resume. \\
    \addlinespace[4pt]

    Interpretive Ambiguity &
    Any indication that users were unsure why the wait was happening or whether the system was still working. \\
    \bottomrule
    \end{tabular}
\end{table}

\end{document}